\documentclass[reprint,aps,prd,nofootinbib,twocolumn,superscriptaddress,preprintnumbers]{revtex4-2}

\usepackage[T1]{fontenc}
\usepackage[usenames,dvipsnames]{xcolor}
\usepackage{aas_macros,graphicx,hyperref,mathrsfs,orcidlink,subfigure}
\hypersetup{colorlinks=true,citecolor=red,linkcolor=red,urlcolor=blue}

\DeclareUnicodeCharacter{2299}{\ensuremath{\odot}}

\usepackage{booktabs}

\usepackage{ulem}

\usepackage{graphicx,amssymb,amsmath,amsthm,amsfonts,epstopdf,epsfig,epsf,times}

\usepackage{amsmath}

\usepackage{esint}

\newcommand{\be}{\begin{equation}}
\newcommand{\ee}{\end{equation}}
\newcommand{\bea}{\setlength\arraycolsep{2pt} \begin{eqnarray}}
\newcommand{\eea}{\end{eqnarray}}

\newcommand{\mP}{{\mathcal P}}

\newcommand{\mR}{{\mathcal R}}

\newcommand{\md}{\mathrm{d}}
\newcommand{\mA}{\mathcal{A}}
\newcommand{\mD}{\mathcal{D}}

\def\D{\Delta}
\def\f{\frac}

\def\t{\theta}

\def\be{\begin{equation}}
\def\ee{\end{equation}}
\def\bag{\begin{aligned}}
\def\eag{\end{aligned}}
\def\bea{\begin{eqnarray}}
\def\eea{\end{eqnarray}}
\def\ba{\begin{array}}
\def\ea{\end{array}}

\def\bc{\begin{center}}
\def\ec{\end{center}}

\begin{document}

\title{Self-lensing of gravitational waves from binary extreme-mass-ratio inspirals: \\ Extended caustics and paired flashes}

\author{Jiewei Huang$^{\star}$}
\affiliation{School of Physics, Peking University, No.5 Yiheyuan Rd, Beijing
100871, P.R. China}

\author{Yifan Ou$^{\star}$}
\affiliation{School of physics and astronomy, Beijing Normal University, Beijing, 100875, P. R. China}

\author{Yehui Hou}
\email{contact: yehuihou@sjtu.edu.cn}
\affiliation{Tsung-Dao Lee Institute, Shanghai Jiao-Tong University, Shanghai, 201210, P. R. China}

\author{Minyong Guo}
\affiliation{School of physics and astronomy, Beijing Normal University, Beijing, 100875, P. R. China}
\affiliation{Key Laboratory of Multiscale Spin Physics (Beijing Normal University), Ministry of Education, Beijing, 100875, China}

\author{Bin Chen}
\affiliation{Institute of Fundamental Physics and Quantum Technology, \& School of Physical Science and Technology, Ningbo University, Ningbo, Zhejiang, 315211, China}
\affiliation{Zhejiang Key Laboratory of Extreme Universe, \& BINGO Center, 
    Ningbo University, Ningbo, Zhejiang 315211, China }
\affiliation{Center for High Energy Physics, \& School of Physics, Peking University, No.5 Yiheyuan Rd, Beijing, 100871, P. R. China}

\begingroup
\renewcommand\thefootnote{$\star$}
\footnotetext{co-first authors}
\endgroup

\begin{abstract}
We investigate strong gravitational lensing of gravitational waves from a compact source orbiting a Kerr black hole, motivated by self-lensing in binary extreme-mass-ratio inspirals. In the geometric-optics limit, we construct a point-to-point lens map from null-geodesic integrals and use the Jacobi map to compute the strain magnification of individual images. While lensed signals are generally demagnified, we show that they can be strongly amplified near the extended caustics of the Kerr lens map. For suitable source--observer geometries, the orbit crosses a caustic twice per revolution, producing two amplification peaks. Their arrival-time separation depends on the black-hole spin, observer inclination, and orbital radius through both the source motion and propagation delays. These signatures might offer a means of constraining the source--observer geometry and probing the spacetime of the central black hole.
\end{abstract}

\maketitle

\section{Introduction}

Black holes produce strong-lensing phenomena beyond the weak-deflection regime, including multiple images, large deflection angles, and nontrivial caustics \cite{Ohanian:1987xga, Perlick:2004tq, Bozza:2010xqn}.
For gravitational waves (GWs), these effects modify the
observed amplitudes, phases, and arrival times
\cite{Dai:2017huk, Li:2019osa, Xu:2021bfn, Wierda:2021upe, Oguri:2022zpn, Grespan:2023cpa, 
Saketh:2025cwf, Barsode:2025agk}.
Binary extreme-mass-ratio inspirals (b-EMRIs), in which a stellar-mass compact binary orbits a supermassive black hole (SMBH), provide a natural setting for strong-field lensing
in hierarchical systems \cite{Chen:2017xbi, Chen:2018axp}.
Such systems may form near migration traps in AGN disks \cite{McKernan:2011gv, Bellovary:2015ifg, Tagawa:2019osr, Peng:2021vzr}.
The inner binary and the outer orbit can emit GWs at widely separated frequencies: the inner-binary signal may lie in the sensitive band of ground-based detectors \cite{Abbott:2016xvh}, whereas radiation from the outer orbit is accessible to space-based observatories
\cite{LISA:2017pwj, TianQin:2015yph, Hu:2017mde, Yu:2021dqx}.
Lensing of the inner-binary radiation by the SMBH can produce multiple images with distinct magnifications, phase shifts, and arrival times.

Previous studies of b-EMRIs have examined orbital dynamics and relativistic GW generation
\cite{Han:2018hby, Fang:2019mui, Yu:2020dlm, Cardoso:2021vjq, Gong:2021kbp, Fang:2021iyf, Camilloni:2023xvf, Yin:2024nyz, Zhang:2024ibf, Santos:2025ass, Santos:2026lzq},
as well as SMBH-induced redshift and lensing and their implications for observed signals and source-parameter inference
\cite{Chen:2017xbi, DOrazio:2019fbq, Zhang:2023cdh, Wu:2026pwl, Martens:2026mku}.
Complementary work on black-hole imaging has characterized image formation and magnification
\cite{Gralla:2019drh, Zhou:2024dbc, Frost:2024pxa, Frost:2025wvh},
while studies of strongly lensed GWs have elucidated image-dependent phase shifts
\cite{Ezquiaga:2020gdt}.
Here, we investigate the geometric magnification of GWs propagating through the SMBH spacetime, focusing on how caustic crossings by an orbiting inner binary determine the strain-envelope amplification and the timing of distinct amplification episodes.

\begin{figure}[htbp]
    \includegraphics[width=1\columnwidth]{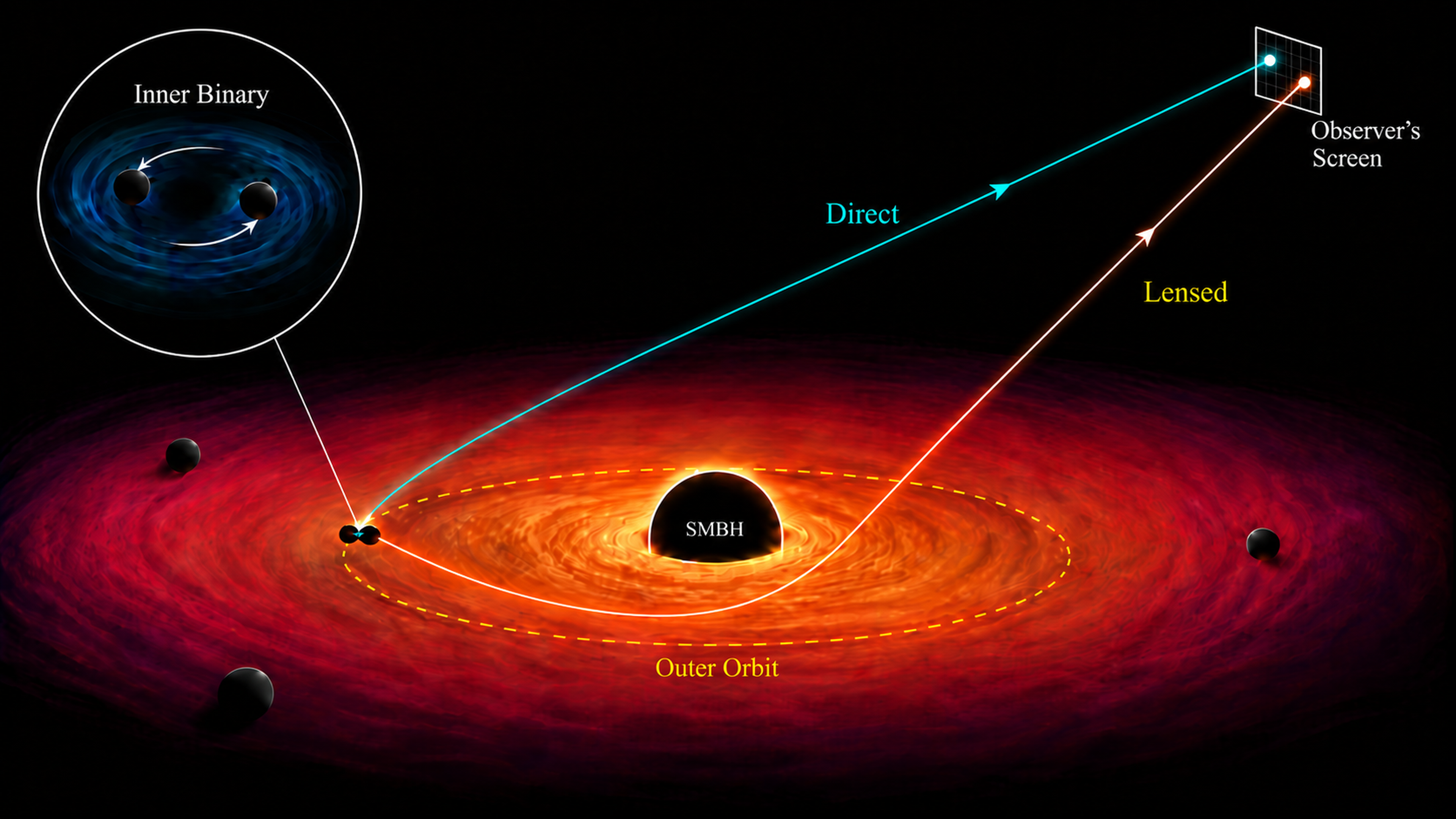}
    \caption{GW propagation from a compact binary orbiting
    an SMBH. The inner binary is treated as a localized
    emitter. Distinct null geodesics connecting the source
    and observer produce direct and lensed signals.}
    \label{fig:sche}
\end{figure}

To isolate propagation effects from the binary's internal evolution, we treat the inner binary as a localized emitter whose center of mass follows a stable circular equatorial orbit around a Kerr black hole.
At leading order in geometric optics, radiation reaches a distant observer along null geodesics associated with a direct image and lensed images
\cite{Luminet:1979nyg, Gralla:2019xty, Gralla:2019drh}, as illustrated in Fig.~\ref{fig:sche}.
Although lensed images are typically faint away from caustics \cite{Bozza:2006nm},
they can be strongly amplified near caustic crossings.
In spherically symmetric spacetimes, lensing caustics are degenerate and aligned with the optical axis; in Kerr spacetime, they can be extended and displaced, with spin-dependent azimuthal drift
\cite{Perlick:2004tq, Sereno:2007gd, Bozza:2008mi}.
An orbiting source can therefore cross distinct caustic folds and produce separate episodes of enhanced lensed radiation.

We solve the point-to-point Kerr lens map to determine the screen positions and propagation times of individual images.
Ray-bundle focusing is computed using bilocal geodesic operators and the associated Jacobi map
\cite{perlick2003ray, Uzun:2018yes, Grasso:2018mei,
Serbenta:2021tzv, Grasso:2021iwq, Korzynski:2024jqt}.
With directional emission normalizations matched and redshift effects treated separately, this construction yields the geometric magnification and the corresponding strain-envelope enhancement.
The point-source geometric-optics magnification diverges at optical caustics; finite source size and wave effects regularize this divergence.
Our calculation thus provides the geometric focusing and timing contributions, rather than a complete observed waveform, which additionally requires the evolving source emission, image-dependent frequency shifts, phase and polarization transport, and detector response.

For the configurations considered here, the direct image remains regular because the source orbit does not intersect the primary caustic tube.
Higher-order images can nevertheless be strongly amplified near folds of the secondary caustic.
If the circular source orbit crosses this caustic at two distinct azimuths and emission persists through both crossings, two amplification episodes, hereafter termed flashes, can occur per outer orbital period.
These flashes originate from distinct emission events at separate caustic crossings, not from multiple images of a single event.
Their arrival-time separation depends on both the orbital interval between crossings and the difference in propagation times, and hence on the black-hole spin, source orbital radius, and observer inclination.
In the large-radius, nearly edge-on regime considered here, the leading time separation scales quadratically with the dimensionless spin and as the source radius to the three-halves power, with radius and time expressed in units of the SMBH mass.
These results connect Kerr caustic geometry and source orbital motion to the amplification and timing of lensed radiation, providing a geometric basis for subsequent detectability assessments.

The remainder of this paper is organized as follows. 
Sec.~\ref{sec:theo} assesses the scale hierarchy underlying the geometric optics, and develops GW amplitude transport, the bilocal geodesic operators, and the Jacobi map. Sec.~\ref{sec:KerrLensing} constructs the point-to-point Kerr lens map. 
Sec.\ref{sec:caustic} examines the geometry of the caustic surfaces. 
Sec.~\ref{sec:magnif} presents GW magnifications and the timing of paired flashes. Sec.~\ref{sec:discussion} summarizes the results, their limitations, and directions for future work.

\section{General MODELING}\label{sec:theo}

\subsection{Order-of-magnitude estimates}\label{sec:oom}

We first compare the GW wavelength of the inner binary with the gravitational radius of the SMBH. For an equal-mass, quasi-circular binary with component mass $m$ and separation $d$, the wavelength at the dominant GW frequency is
\begin{equation}
\frac{\lambda_{\rm GW}}{R_g}
\approx 6 \times 10^{-3}
\left(\frac{m}{30M_\odot}\right)
\left(\frac{10^6 M_\odot}{M}\right)
\left(\frac{d}{10\,r_g}\right)^{3/2}\!\!,
\label{eq:lambda_over_Rg}
\end{equation}
where $r_g=2Gmc^{-2}$ and $R_g=GMc^{-2}$. Thus, for $m \sim 10M_\odot$, $M\sim 10^6M_\odot$ and $d=10$--$10^2r_g$, $\lambda_{\rm GW}\ll R_g$, supporting a geometric-optics treatment of propagation through the SMBH's strong-field region~\cite{Zhang:2023cdh}. This estimate is specific to the inner-binary frequency; it does not establish geometric-optics validity across an entire observational frequency band.

We next compare the inner-binary inspiral time with the outer orbital period. For an isolated equal-mass circular binary, the leading-order time to coalescence is~
$t_{\rm GW}\simeq 5c^5d^4(512G^3m^3)^{-1}$ \cite{Peters:1964zz}.
Taking the Schwarzschild ISCO period, $T_{\mathrm{ISCO}}=2\pi 6^{3/2}GMc^{-3}$, as a reference outer-orbital timescale gives
\begin{equation}
\frac{t_{\rm GW}}{T_{\mathrm{ISCO}}}
\approx 5\times10^{-4}
\left(\frac{m}{30M_\odot}\right)
\left(\frac{10^6 M_\odot}{M}\right)
\left(\frac{d}{10\,r_g}\right)^4\!.
\label{eq:tc_over_Tisco}
\end{equation}
The period of an actual Kerr outer orbit depends on its radius and the SMBH spin. The isolated-binary estimate is appropriate at order-of-magnitude level provided that the SMBH tide does not substantially alter the inner binary's evolution.

For $m=30M_\odot$, $M=10^6M_\odot$, and $d=10$--$10^2r_g$, Eq.~\eqref{eq:tc_over_Tisco} gives $t_{\rm GW}/T_{\mathrm{ISCO}}\sim5\times10^{-4}$ -- $5$. The same lens map applies in both limits: at small separations, the short-lived signal samples an approximately fixed source position; at larger separations, the source may traverse much of its outer orbit, tracing image tracks on the observer's sky \cite{DOrazio:2019fbq}. We approximate the inner binary as a point source on the scale of the outer orbit and neglect three-body effects such as Lidov--Kozai oscillations \cite{Antonini:2012ad, Hoang:2017fvh, Liu:2019tqr}. Unless stated otherwise, we use geometric units, $G=c=1$.



\subsection{WKB limit}\label{sec:WKBlimit}

We consider linear metric perturbations, $g_{\mu\nu}+h_{\mu\nu}$, about a vacuum background $g_{\mu\nu}$. Indices are raised and lowered with the background metric, and $\nabla_\mu$ denotes its covariant derivative. In the Lorenz gauge,
$\nabla_\mu(h^{\mu\nu}-\tfrac{1}{2}g^{\mu\nu}h)=0$, with $h=g^{\mu\nu}h_{\mu\nu}$, the vacuum linearized Einstein equation reads
\bea\label{eq:eom}
\nabla^{\alpha}\nabla_{\alpha} h_{\mu\nu}
-2R_{\mu\alpha\nu\beta}h^{\alpha\beta}=0 \,,
\eea
where the background Riemann tensor is defined by
$[\nabla_\alpha,\nabla_\beta]v^\mu
=-R^\mu{}_{\nu\alpha\beta}v^\nu$ with $v$ the arbitrary vector.

In the geometric-optics limit, the GW wavelength is much shorter than the characteristic scale of background variation. When this scale is set by the lens radius, the condition becomes $\lambda_{\rm GW}\ll R_g$. We adopt the WKB ansatz
\bea
h_{\mu\nu}=\mathrm{Re}\!\left[\mA_{\mu\nu}(x)\,e^{iS(x)/\epsilon}\right] \,,
\eea
where $\mA_{\mu\nu}$ is a slowly varying complex amplitude and $\epsilon\ll1$ is a bookkeeping parameter encoding the separation between the wavelength and the background scale \cite{isaacson1968gravitational}.

Defining $k_\mu=\nabla_\mu S$, the leading, $\mathcal{O}(\epsilon^{-2})$, terms in Eq.~\eqref{eq:eom} give the eikonal equation $k_\mu k^\mu=0$. Since $k_\mu$ is a gradient, this also implies $k^\nu\nabla_\nu k^\mu=0$, i.e., the rays are affinely parameterized null geodesics. At leading WKB order, the Lorenz condition gives
$k_\mu\mA^{\mu\nu}-\tfrac{1}{2}k^\nu\mA^\alpha{}\alpha=0$.
The residual gauge freedom can be used locally to choose a transverse, traceless representative of the radiative amplitude, satisfying
$k\mu\mA^{\mu\nu}=0$ and $\mA^\mu{}_\mu=0$.

Writing $\mA^{\mu\nu}=\mA\,e^{\mu\nu}$, where $\mA$ is a scalar amplitude and $e^{\mu\nu}\bar e_{\mu\nu}=1$, the $\mathcal{O}(\epsilon^{-1})$ field equation yields
\bea\label{eq:Ae01}
\left(k^\alpha\!\nabla_\alpha+\tfrac{1}{2}\nabla_\alpha k^\alpha \right)\mA=0 \,, \quad k^\mu\nabla_\mu e^{\alpha\beta}=0 \,.
\eea
Thus, the polarization tensor is parallel transported, while the amplitude obeys the flux-conservation law
$\nabla_\mu(\mA^2 k^\mu)=0$.
Along a ray with affine parameter $\lambda$, chosen such that $k^\mu=\md x^\mu/\md\lambda$, one has $k^\mu\nabla_\mu=\md/\md\lambda$ when acting on scalars. The amplitude therefore evolves as
\bea\label{eq:Aint}
\begin{aligned}
\mA(\lambda) =
\mA(\lambda_0)
\exp\!\left[
-\frac{1}{2}\int_{\lambda_0}^{\lambda}
\md\lambda'\,\nabla_\alpha k^\alpha(\lambda')
\right]
\,.
\end{aligned}
\eea
Here $\nabla_\mu k^\mu$ describes the fractional rate of change of the ray bundle's cross-sectional area, so that $\mA$ scales as the inverse square root of that area \cite{Flanagan:2005yc}. For a spherically diverging bundle in flat spacetime, this reduces to $\mA\propto d^{-1}$, where $d$ is the distance from the source. In curved spacetime, gravitational focusing and defocusing modify this evolution. 

In vacuum geometric optics, electromagnetic and gravitational waves both follow null geodesics and obey an analogous amplitude-flux conservation law. The distinction relevant here is primarily observational rather than dynamical: GW measurements retain coherent amplitude and phase information, whereas electromagnetic observations are often modeled using time-averaged intensities or spectral radiative-transfer quantities \cite{Lindquist:1966igj, gammie2012formalism}.

\subsection{Strain envelope}\label{sec:bundle}

Black-hole lensing changes the GW strain envelope through the focusing and defocusing of null geodesic bundles. In geometric optics, the amplitude transport law in Eq.\eqref{eq:Aint} conserves $\mathcal{A}^2\md\mathscr{A}$ along each regular bundle, where $\md\mathscr{A}$ is its physical transverse area. The lensing amplification is therefore determined by the evolution of this area, which we describe covariantly using the Jacobi matrix \cite{perlick2003ray,Harte:2012uw}. Details of the screen geometry and amplitude normalization are given in Appendix\ref{app:JacobiFlux}.

Let $k^\mu$ be an affinely parametrized null tangent, $\xi^\mu$ a connecting vector between neighboring geodesics, and $\ell^\mu=\mathrm{D}_{\lambda}\xi^\mu$, with $\mathrm{D}_{\lambda}=k^\alpha\nabla_\alpha$. The bilocal geodesic operators (BGOs) propagate these deviations according to \cite{Grasso:2018mei,Serbenta:2021tzv,Grasso:2021iwq,Korzynski:2024jqt}
\bea\label{BGO01}
\begin{pmatrix}
\xi^\mu(\lambda)\\[1mm]
\ell^\mu(\lambda)
\end{pmatrix}
=
\begin{pmatrix}
W_{XX}^{\mu}{}_{\nu'} & W_{XK}^{\mu}{}_{\nu'}\\[1mm]
W_{KX}^{\mu}{}_{\nu'} & W_{KK}^{\mu}{}_{\nu'}
\end{pmatrix}
\begin{pmatrix}
\xi^{\nu'}(\lambda_s)\\[1mm]
\ell^{\nu'}(\lambda_s)
\end{pmatrix}\,,
\eea
where primed indices refer to the source, all blocks are evaluated at $(\lambda,\lambda_s)$, and the full propagator satisfies $W(\lambda_s,\lambda_s)=\mathbb{I}_{8}$. For a source-vertex bundle, $\xi^\mu(\lambda_s)=0$, so $W_{XK}$ maps initial tangent perturbations to ray separations. Projection onto a parallel-transported orthonormal screen basis $\{e_A{}^\mu\}_{A=1,2}$ defines the Jacobi matrix,
\bea\label{eq:JacobiBGO}
\mD^A{}_B(\lambda,\lambda_s)
=
e^A{}_\mu(\lambda)\,
W_{XK}^{\mu}{}_{\nu'}(\lambda,\lambda_s)\,
e_B{}^{\nu'}(\lambda_s)\,.
\eea
With our curvature convention, it obeys
\bea\label{eq:JacobiEvolution}
\mathrm{D}_{\lambda}^2\,\mD^A{}_B
=
R^\mu{}_{\alpha\nu\beta}
\,e^A{}_\mu e_C{}^\nu k^\alpha k^\beta\,\mD^C{}_B\,,
\eea
with $\mD(\lambda_s,\lambda_s)=0$ and
$\mathrm{D}_{\lambda}\mD(\lambda_s,\lambda_s)=\mathbb{I}_{2}$.
Because the screen basis is parallel transported, $\mathrm{D}_{\lambda}$ acts on screen components as an ordinary affine-parameter derivative.

Choosing the initial screen in the source rest frame, we define $\omega_s=-u_s\cdot k_s$. An angular displacement $\md\vartheta^B$ at fixed $\omega_s$ corresponds to
$\delta k_s^\mu=\omega_s e_B{}^\mu\md\vartheta^B$, giving
\bea\label{eq:JacobiDisplacement}
\begin{aligned}
&\xi_o^A
=
\omega_s\mD^A{}_B(\lambda_o,\lambda_s)\md\vartheta^B\,, \\
&\md\mathscr{A}_o
=
\omega_s^2
|\det\mD(\lambda_o,\lambda_s)|\,\md\Omega_s\,.
\end{aligned}
\eea
Thus, $\omega_s^2|\det\mD|$ is the physical transverse area at the observer per unit source-frame solid angle. This combination is invariant under affine rescaling and includes the dependence of the angular map on the source motion. The quantity $\omega_s$ fixes the tangent normalization; it is the physical source-frame angular frequency when $k^\mu$ is the wave four-vector.

The strain envelope follows from
\be\label{eq:GWAmplitudeJacobi}
\mathcal{A}^2\,\md\mathscr{A}
=\mathrm{constant}\,,
\quad
\mathcal{A}\propto|\det\mD|^{-1/2}\,,
\ee
where the proportionality holds along a regular bundle with fixed emission data and tangent normalization. The emitted amplitude is specified on a finite matching section in the radiative wave zone, rather than at the idealized bundle vertex. For lensed and reference propagation with the same emitted amplitude normalization per unit source-frame solid angle, we define
\bea\label{eq:GeometricMagnification}
\mu=
\frac{\omega_{s,\rm ref}^{2}|\det\mD_{\rm ref}|}
     {\omega_s^2|\det\mD|}\,,
\quad
\frac{\mathcal{A}_o}{\mathcal{A}_{o,\rm ref}}
=
\sqrt{\mu}\,.
\eea
This definition isolates the geometric amplification from intrinsic variations of the emission with time, direction, or frequency. 
We further adopt a spherical-dilution reference,
$\omega_{s,\rm ref}^{2}|\det\mD_{\rm ref}|=r_o^2$,
for which
\be\label{eq:AngularMagnification}
\mu=
\frac{r_o^2}{\omega_s^2|\det\mD|}
\simeq \frac{\md\Omega_s}{\md\Omega_{\rm o}}\,.
\ee
The last relation holds in an asymptotically flat receiving region with nearly normal incidence, where
$\md\mathscr{A}_o\simeq r_o^2\md\Omega_{\rm o}$.
Here $\md\Omega_{\rm o}$ is the solid angle subtended by the bundle footprint at the center of the receiving sphere, rather than an image solid angle on the observer's sky.

For multiple propagation paths, the observed strain is obtained by coherently summing the transported waveforms in a common observer-frame polarization basis before applying the detector response. 
Each contribution carries its retarded emission data, propagation delay and phase. At a caustic away from the source vertex, $\det\mD=0$, and the geometric-optics amplitude diverges. This singularity signals the breakdown of geometric optics, not an infinite physical strain; a wave-optical treatment is required near the caustic \cite{Harte:2012uw}.

\begin{figure*}[htbp]
    \includegraphics[width=1.7\columnwidth]{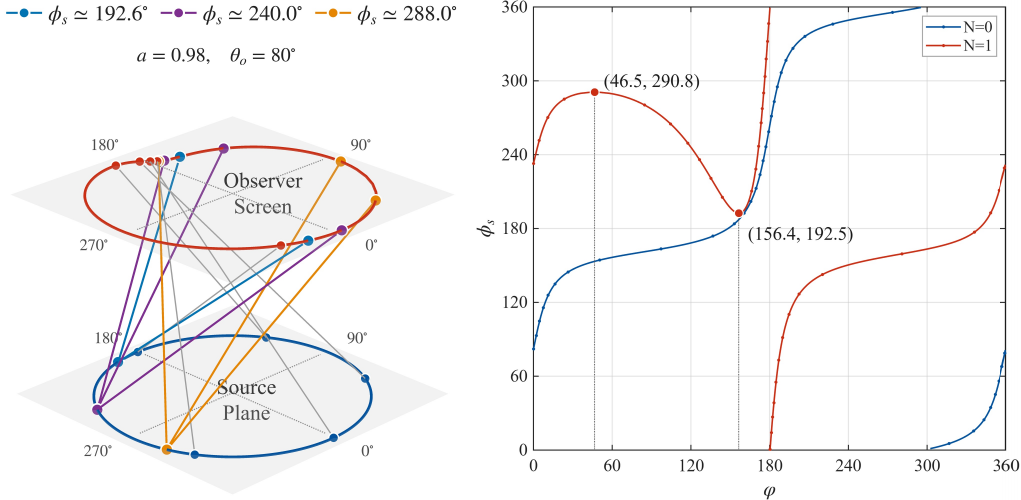}
    \caption{\textbf{Left:} Lensed image loci on the observer's screen for an equatorial source ring at $r_s=4$, showing the correspondence between screen polar angle $\varphi$ and source azimuth $\phi_s$. \textbf{Right:} The mappings $\varphi(\phi_s)$ for direct and lensed images of the same ring. In both panels, $a=0.98$ and the observer is located at $\theta_o=80^\circ$ and $\phi_o=0^\circ$.}
    \label{fig:map}
\end{figure*}

\section{Kerr lensing}\label{sec:KerrLensing}

Now we specify to the Kerr spacetime, characterized by the black-hole mass $M$ and angular momentum $J$~\cite{kerr1963gravitational}. In Boyer--Lindquist coordinates $(t,r,\theta,\phi)$, the conserved quantities $l=L_z/E$ and $\eta=Q/E^2$ specify the axial impact parameter and rescaled Carter constant along geodesics \cite{Carter:1968rr}. Normalizing the photon energy to $E=1$, the wavevector reads $k_{\mu}=(-1,v_r\Delta^{-1}\sqrt{R},v_\theta\sqrt{\Theta},l)$, with
\bea\label{eq:Kerrnulleq}
\begin{aligned}
&R(r) = \mP^2 - \Delta \left[ \eta + \left(l-a\right)^2 \right],\\
&\Theta(\theta) = \eta + a^2\cos^2{\theta} - l^2\cot^2{\theta}.
\end{aligned}
\eea
Here $a=J/M$ is the spin parameter, $\Delta=r^2-2Mr+a^2$, $\mP=r^2+a^2-la$, and $v_r,v_\theta=\pm1$ denote the radial and polar propagation directions outside the event horizon. Integrating the null geodesic equations, the path connecting a source event $(t_s,r_s,\theta_s,\phi_s)$ to an observer event $(t_o,r_o,\theta_o,\phi_o)$ is determined by \cite{Vazquez:2003zm,Beckwith:2004ae,Kraniotis:2010gx}
\bea\label{eq:lenseqs}
\begin{aligned}
& \fint^{r_o}_{r_s} \frac{\md r}{v_r\sqrt{R}}
= \fint^{\theta_o}_{\theta_s} \frac{\md \theta}{v_\theta\sqrt{\Theta}} = \mathcal{T}\,,\\
& \D\phi
= a \fint^{r_o}_{r_s} \frac{2Mr-al}{v_r\Delta\sqrt{R}} \md r
+ l \fint^{\theta_o}_{\theta_s} \frac{\csc^2{\theta}}{v_\theta\sqrt{\Theta}} \md\theta\,,\\
& \D t
= \fint^{r_o}_{r_s} \frac{r^2\Delta+2Mr\mP}{v_r\Delta\sqrt{R}} \md r
+ a^2 \fint^{\theta_o}_{\theta_s} \frac{\cos^2{\theta}}{v_\theta\sqrt{\Theta}} \md\theta\,,
\end{aligned}
\eea
where $\mathcal{T}$ is the ``Mino time" \cite{Mino:2003yg}, a useful parameter to characterize the length of the geodesics; $\D\phi=\phi_o-\phi_s$, $\D t=t_o-t_s$ represent the azimuthal change and propagation time. The integral $\fint$ denotes that $r$, $\t$ are not monotonous along the path, with $v_r$ and $v_\theta$ changing sign at the turning points. 
Eq.~\eqref{eq:lenseqs} can be expressed in terms of elliptic integrals, with appropriate branches determined by turning point features \cite{Gralla:2019ceu}.
It facilities the point-to-point map, well applying to point-source emissions~\cite{Dexter:2009fg, Cardenas-Avendano:2022csp, Zhou:2024dbc}. In what follows, we set the black hole mass $M=1$ for simplicity.

Different null geodesics connecting the same emission position to the observer result in multiple images, with different arrival directions and travel times. For an equatorial source, we can label each trajectory by the equatorial crossing number $N$ after emission, with $N=0,1,2,\dots$ corresponding to direct, lensed, second-order, and subsequent images. For a northern-hemisphere observer with $0<\theta_o<\pi/2$, $N$ is related to the number of polar turning points $m$ by
\bea\label{eq:mN}
N=m-\frac{v_{\theta_o}+1}{2},
\quad m=0,1,2,\dots,
\eea
where $v_{\t_o} = \text{sign}(k_\theta)$ at the observer, and only combinations satisfying $N\geq0$ are observable. At sufficiently high order, the trajectories approach unstable photon orbits, and their images accumulate near the critical curve~\cite{Bardeen:1972fi, Luminet:1979nyg}.

To construct the lens map from equatorial sources to a distant observer, we decompose the equatorial plane into series of rings with radius $r_s$.
Inverting the first equation in Eq.~\eqref{eq:lenseqs} gives the transfer function $r_s=r(N,l,\eta,\theta_o)$ that helps us to select the impact parameter range under fixed $r_s, N, \t_o$ \cite{Li:2004aq, Gralla:2019drh}. Then, the number of raidal turning points, labeled by $w$, can be easily determined by $w=\mathrm{H}\big(\mathcal{T}-\int^{r_o}_{r_s}\md r R^{-1/2}\big)$,
where $\mathrm{H}$ is the Heaviside function, with $\mathrm{H}(0)=0$ for the outward unturned branch. 
In Kerr spacetime, the radial motion has either zero or one turning point for escaping rays \cite{Gralla:2019ceu}.
The azimuthal and temporal equations then determine the source azimuth and travel time along each trajectory; travel-time differences yield the inter-image delays.

At the observer's screen, we adopt the screen basis
$e_{\alpha}^{\mu}=(r_o\sin\theta_o)^{-1}(\partial_\phi)^\mu$
and
$e_{\beta}^{\mu}=-r_o^{-1}(\partial_\theta)^\mu$. According to Bardeen's convension, 
the standard screen coordinates are given by \cite{Bardeen:1972fi}
\bea
\alpha=-\frac{l}{\sin\theta_o}\,,
\quad
\beta=v_{\theta_o}\sqrt{\Theta(\theta_o)}\,.
\eea
We further introduce screen polar coordinates $b=\sqrt{\alpha^2+\beta^2}$ and $\varphi=\operatorname{atan2}(\beta,\alpha)$ to parametrize the image loci. Hereafter, we fix the observer at $\phi_o=0^\circ$.

\begin{figure*}[htbp]
	\includegraphics[width=2\columnwidth]{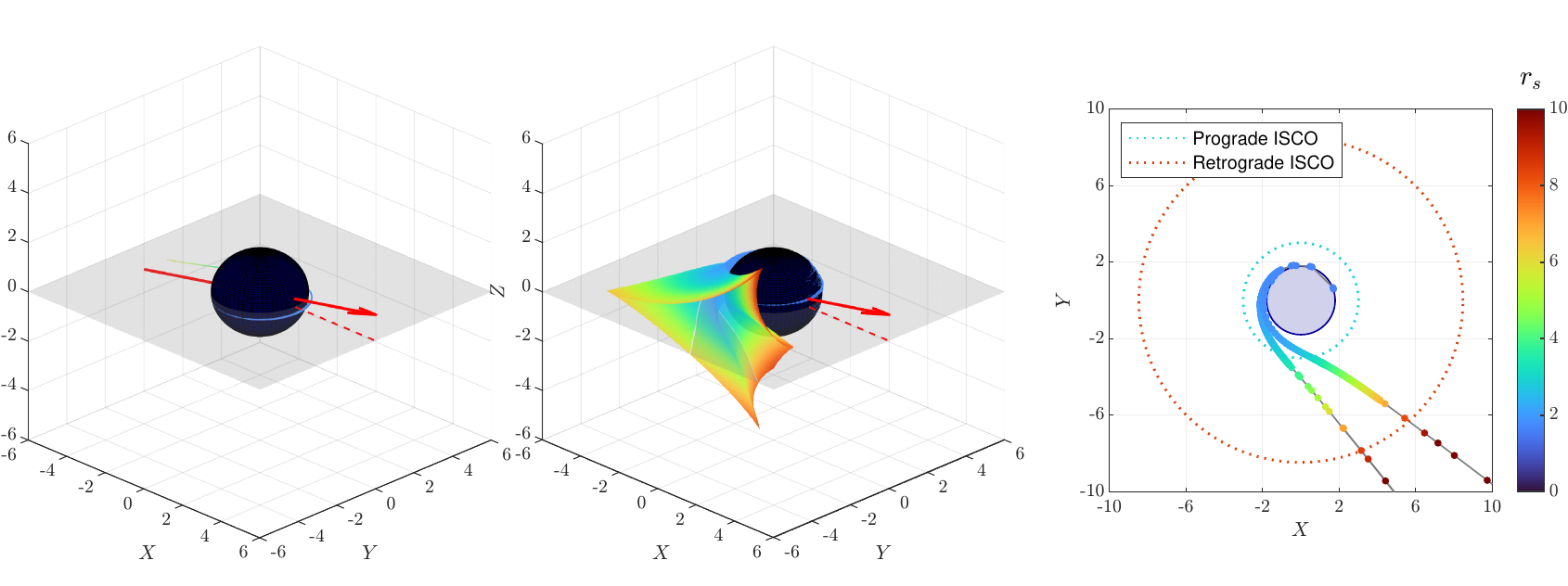}
	\includegraphics[width=2\columnwidth]{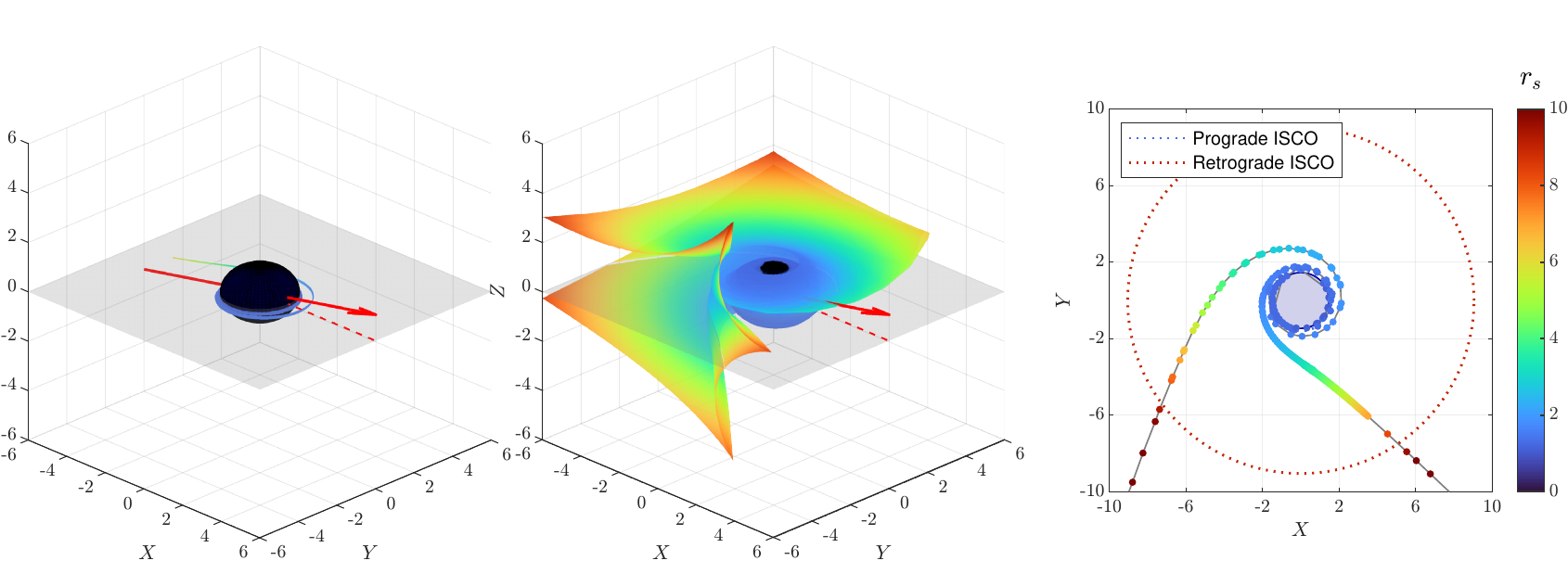}
\caption{Caustic surfaces for $m=1$ (\textbf{left}) and $m=2$ (\textbf{middle}) around Kerr black holes with spins $a=0.8$ (\textbf{top}) and $a=0.998$ (\textbf{bottom}), viewed at an inclination $\t_o=80^\circ$. The surfaces are shown in Cartesian Kerr--Schild coordinates $(X,Y,Z)$, with color indicating the radial coordinate. The red arrow points toward the observer; the dashed line shows its projection onto the equatorial plane (gray). The \textbf{right} column shows the $m=2$ caustics on the equatorial plane.}
	\label{fig:3DCS}
\end{figure*}

\begin{figure*}[htbp]
\centering
\hspace*{-7mm}
    \includegraphics[width=2\columnwidth]{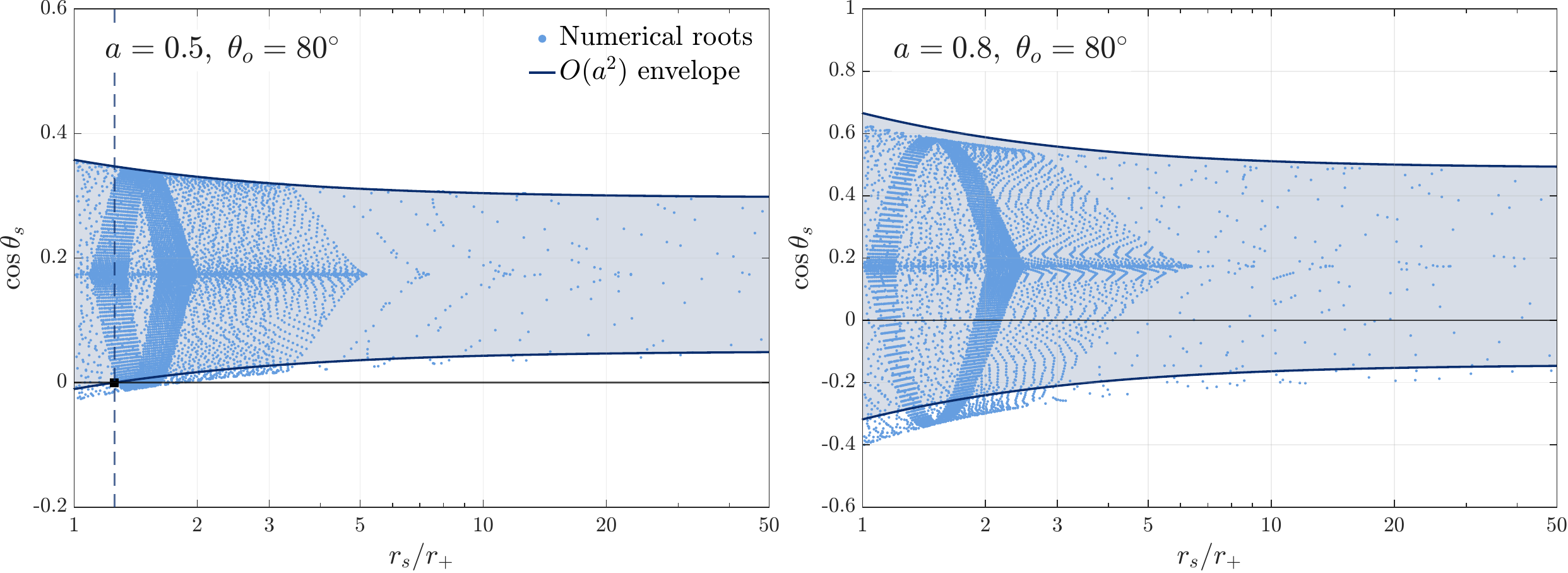}
\caption{Secondary caustics in the $(r_s,\t_s)$ plane, where $r_+=1+\sqrt{1-a^2}$ is the event-horizon radius. The shaded band shows Bozza's approximation through $\mathcal{O}(a^2)$, and the blue dots show caustic points computed numerically using the BGO method. The horizontal line marks the equatorial plane; the vertical dashed line marks the radius at which it intersects the boundary of the $\mathcal{O}(a^2)$ approximation. The nonuniform distribution of blue dots reflects the sampling induced by the parameterization of the CS on the observer's sky.}
    \label{fig:bozzacompare}
\end{figure*}

\begin{figure}[htbp]
\centering
	\includegraphics[width=0.75\columnwidth]{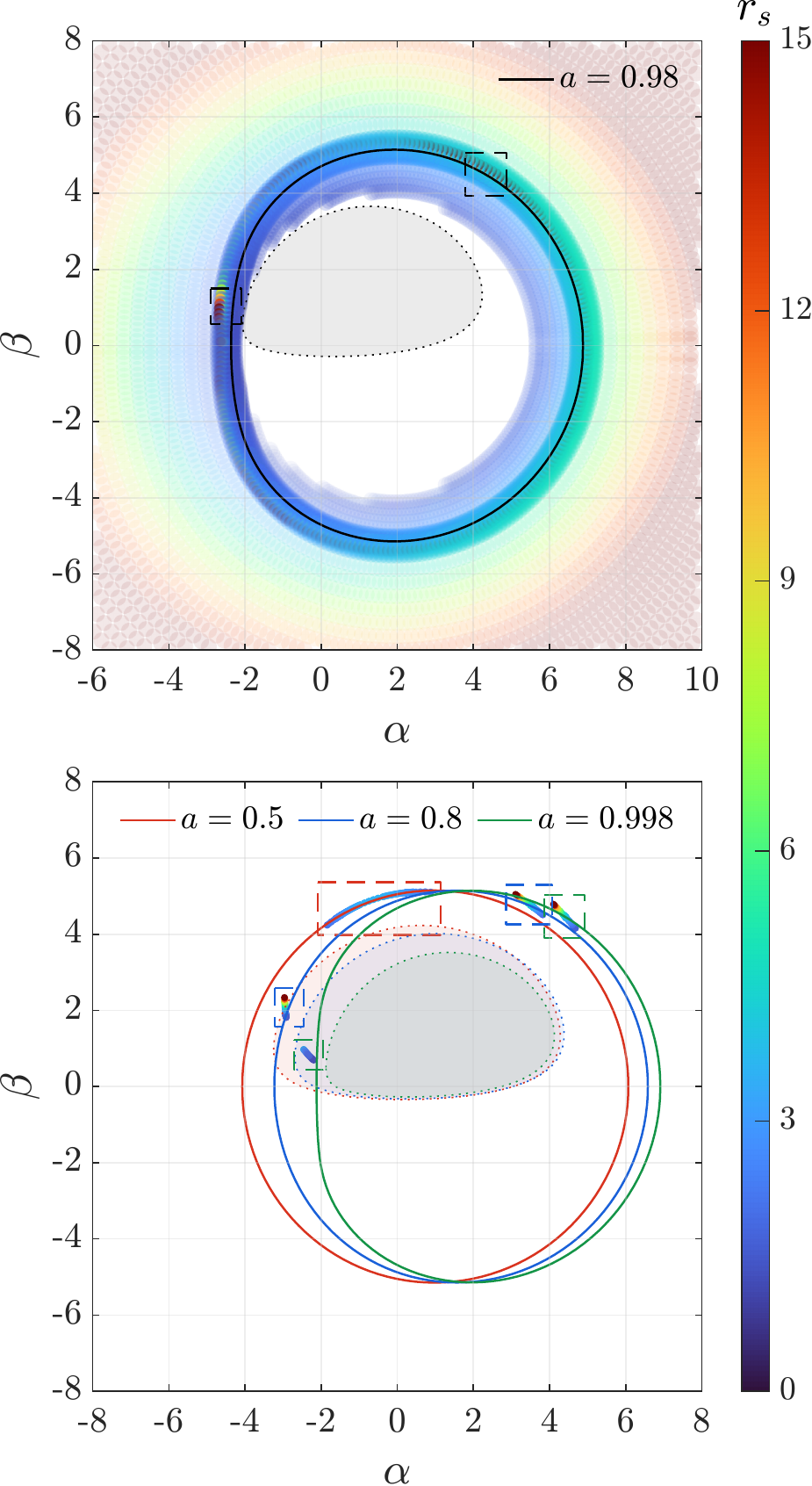} 
\caption{Images of caustics on the observer's screen for an inclination of $\theta_o = 80^\circ$. \textbf{Top:} Image-plane positions for different source radii and latitudes. Colors indicate the radial coordinate, while progressively lighter shades denote latitudes increasing from the equatorial plane toward the north pole for $m = 1,2$. Smaller dots near the critical curve mark higher-order images with larger $m$. \textbf{Bottom:} Caustic images for equatorial sources with $m = 2$, $v_{\t_o} = 1$, and $N = 1$, tightly clustered near the critical curve. The shaded region denotes the projected horizon on the screen for size reference.}
	\label{fig:CSimage}
\end{figure}

Fig.~\ref{fig:map} shows the source-azimuth maps $\phi_s(\varphi)$ along the direct and lensed image loci. The direct-image locus has a characteristic hat-like shape, and its source-azimuth map is monotonic. The lensed-image locus is strongly distorted and lies close to the critical curve associated with unstable spherical photon orbits \cite{Bardeen:1972fi}, and at sufficiently high observer inclination, $\phi_s(\varphi)$ becomes nonmonotonic. For $\theta_o = 80^\circ$ (right panel of Fig.~\ref{fig:map}), the map has two local extrema at $(\phi_s,\varphi) = (192.5^\circ,156.4^\circ)$ and $(290.8^\circ,46.5^\circ)$. There is one image of the specified order for $0^\circ < \phi_s < 192.5^\circ$ and $290.8^\circ < \phi_s < 360^\circ$, but three for $192.5^\circ < \phi_s < 290.8^\circ$.\footnote{Similar image-multiplicity patterns have been investigated in Kerr--Newman spacetime \cite{Hou:2022gge}; the number of images depends on the caustic geometry and the source trajectory.}

These changes in image multiplicity occur when the source crosses a caustic of the lens map \cite{Perlick:2004tq}. At a generic fold crossing, two images merge on the corresponding image-plane critical curve, and their point-source magnifications diverge in geometric optics \cite{Matsunaga:2006uc}. A transverse intersection of the source ring with a fold appears as a local extremum of $\phi_s(\varphi)$ along the $N=1$ image locus. The two extrema thus delimit the three-image interval: an image pair is created at one crossing and annihilated at the other, while a third image persists.

\section{Caustic structures}\label{sec:caustic}

Caustics occur where the Jacobi map becomes singular, $\det\mD=0$, marking conjugate points along the null geodesics and formally divergent point-source magnifications in geometric optics. For a fixed observer, the spatial loci of these points on the past light cone define caustic surfaces (CSs) \cite{Sereno:2007gd, Bozza:2008mi}. In Kerr spacetime, their geometry depends on the black-hole spin and the observer position.

The CSs form distinct branches labeled by the number $m$ of polar turning points along the connecting null geodesics. Fig.~\ref{fig:3DCS} shows the $m=1$ and $m=2$ branches for an observer at $\t_o=80^\circ$. Although $m$ and the equatorial-crossing number $N$ are related by Eq.~\eqref{eq:mN}, they organize different aspects of the lens map: $N$ labels the image contours at fixed source radius $r_s$, whereas $m$ is used to track the caustic branches as $r_s$ varies.

As shown in the left panel of Fig.~\ref{fig:3DCS}, the primary ($m=1$) caustic is a narrow tube that extends, at large distances from the black hole, approximately along the continuation of the observer--black-hole axis behind the horizon. In a spherically symmetric spacetime, the caustic degenerates to a one-dimensional locus on this axis: exact source--lens--observer alignment then produces an Einstein ring \cite{Virbhadra:1999nm}. For an off-axis observer of a Kerr black hole, rotation breaks this degeneracy. The primary caustic is displaced from the optical axis and develops a finite transverse extent, with a four-cusped astroid-like cross section \cite{Sereno:2007gd}. \footnote{Under the large $r_s$ expansion, the primary caustic has an analytical expression, which we refer to \cite{Sereno:2007gd}.} Closer to the horizon, strong frame dragging causes the tube to wind tightly around the black hole, such that its extension becomes predominantly azimuthal.

The secondary ($m=2$) caustic, which is associated with more strongly deflected rays, is substantially more distorted, as illustrated in the middle panel of Fig.~\ref{fig:3DCS}. It begins to wind at larger radii and possesses an expanding astroid-like cross section, resulting in a considerably larger spatial extent than the primary caustic. This enlarged extent increases the geometrical probability that a source intersects the CS. It does not, however, directly imply a higher detection probability, which additionally depends on the GW energy flux, source population, and instrumental sensitivity. At high spins, the geometry of the secondary CS becomes intricate: its cross section may span one or more azimuthal periods. Such wrapping reflects the strong frame dragging on the relevant geodesic families.

For the adopted ray labeling and observer configuration, caustics with odd $m$ are located predominantly behind the black hole, whereas those with even $m$ lie predominantly in front of it. This alternating arrangement originates from the different winding and polar-turning histories of the associated null geodesics. The handedness of the azimuthal displacement is set by the black-hole spin through frame dragging and reverses under reversal of the spin orientation.

At fixed observer inclination $\t_o$, the equatorial intersection of the secondary CS changes topology as the spin $a$ increases. At low spin, the CS does not intersect the equatorial plane. Once $a$ exceeds $a_0(\t_o)$, a connected intersection appears over a finite range of radii. At $a=a_c(\t_o)$, this intersection changes topology; for $a>a_c(\t_o)$, it consists of two disconnected branches that extend to arbitrarily large radii and become asymptotically straight (right column of Fig.~\ref{fig:3DCS}). These branches arise where the equatorial plane cuts through the astroid-like caustic tube. The number of connected components is
\bea\label{eq:Ncomp}
\mathcal{N}_{\rm comp}\,=\,
\begin{cases}
\,\,0\,, & \,\,a<a_0(\t_o)\,,\\
\,\,1\,, & \,\,a_0(\t_o) < a<a_c(\t_o)\,,\\
\,\,2\,, & \,\,a>a_c(\t_o)\,.
\end{cases}
\eea
For $\t_o=80^\circ$, the transition spins are $a_0\approx0.48$ and $a_c\approx0.56$. These counts describe the topology of the equatorial CS intersection, not the number of crossings by a source on a particular orbit. Away from tangencies, a circular equatorial orbit either misses the CS or crosses it at two distinct azimuths per revolution, entering and then leaving the caustic tube. In the connected regime, such crossings are restricted to a finite range of orbital radii; in the disconnected regime, they can persist to large radii.

Bozza \cite{Bozza:2008mi} derived an analytic approximation for the shape of CSs of arbitrary order using a small-spin expansion. In Fig.~\ref{fig:bozzacompare}, we compare this approximation with our numerical result for the secondary CS in the $(r_s,\t_s)$ plane. The agreement is good throughout the plotted region for $a=0.5$. For $a=0.8$, the approximation captures the overall shape but deviates near the lower $\t_s$ boundary, where higher-order spin corrections may become important. At large $r_s$, the boundaries approach lines of constant $\t_s$, consistent with the asymptotically fixed angular extent of the astroid-like caustic tube.

Fig.~\ref{fig:CSimage} further maps caustic crossings at different source positions $(r_s,\t_s,\phi_s)$ to their image positions on the screen of an observer at $\t_o=80^\circ$, for caustic surfaces with $m\leq2$. Many images associated with the secondary caustic lie along narrow segments near the critical curve (darker colors), where the corresponding rays pass close to unstable bound photon orbits. For equatorial sources at high spin ($a\gtrsim0.8$), these images cluster on the left and right sides of the critical curve (dashed squares), corresponding to the two asymptotic caustic arms in Fig.~\ref{fig:3DCS}. At $a=0.5$, the equatorial caustic-crossing images instead lie predominantly near the upper part of the screen.

The off-equatorial crossings exhibit a different image distribution. Images associated with the narrow $m=1$ caustic tube near
\bea
\t_s^{\rm(tube)} \,\to\, 180^\circ-\t_o
\eea 
trace an approximately ring-shaped outer region (light colors in the upper panel of Fig.~\ref{fig:CSimage}). This region moves away from the critical curve as $r_s$ increases and can span a substantial fraction of the screen. Thus, the narrow angular extent of the caustic tube does not imply a compact distribution of image positions. The ringlike pattern is formed by images of sources at different positions, rather than by an Einstein ring of any individual point source.

\begin{figure*}[htbp]
\centering
    \includegraphics[width=2.1\columnwidth]{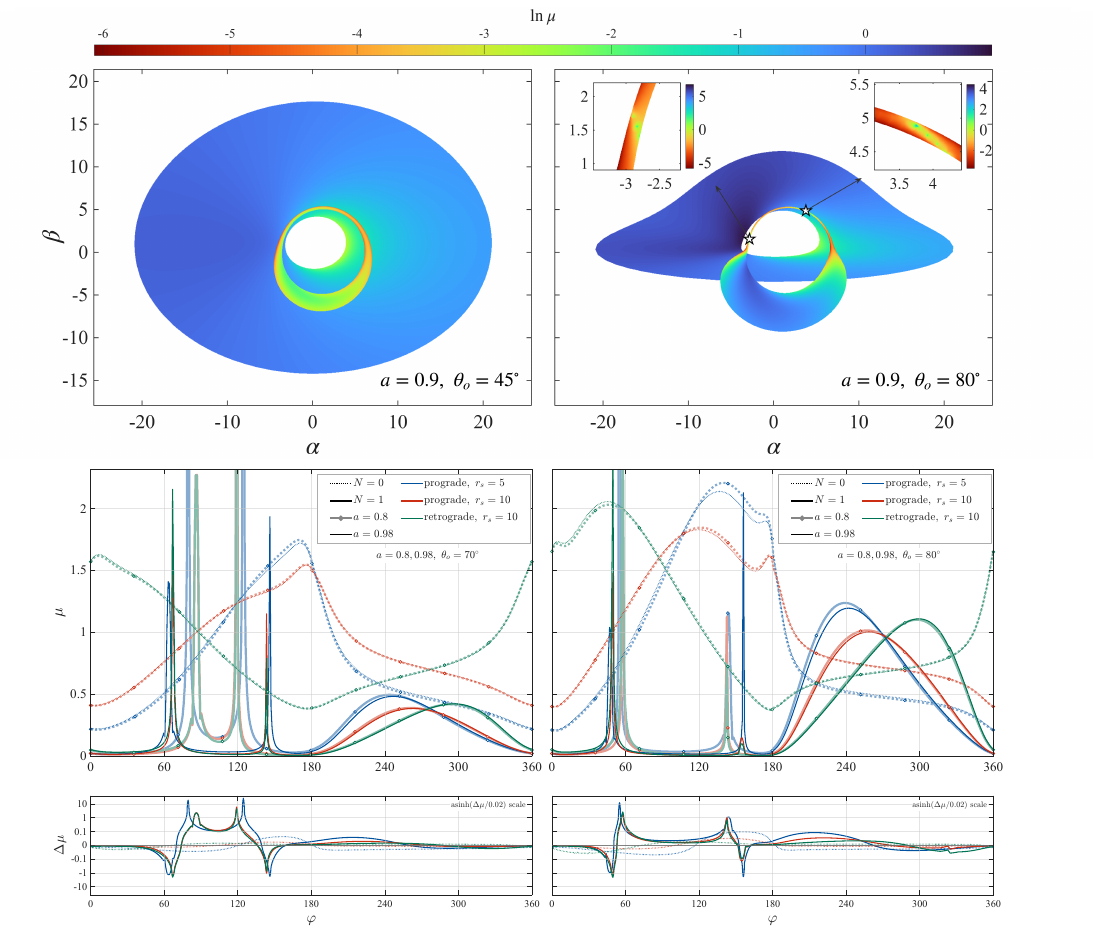}
\caption{
\textbf{Top:} Magnification of the squared GW strain envelope for circular b-EMRIs at different source radii, plotted at the corresponding image positions on the observer's screen. The innermost source radii is given by the ISCO.
\textbf{Middle:} Magnification as a function of screen polar angle $\varphi$ for circular source orbits around Kerr black holes with $a=0.8$ and $0.98$. The observer inclinations are $\t_o=70^\circ$ (\textbf{left}) and $80^\circ$ (\textbf{right}). Prograde orbits are shown at $r_s=5$ and $10$, and retrograde orbits at $r_s=10$.
\textbf{Bottom:} Difference between the magnifications for the two spins shown in the middle row.
}
\label{fig:mumap}
\end{figure*}

Away from higher-order singularities, CSs consist of fold sheets meeting along cusp curves; an astroid-like cross section comprises fold segments connected at cusp points \cite{Perlick:2004tq}. As a point source approaches a fold from the side with an additional image pair, the two images coalesce on the critical curve \cite{Matsunaga:2006uc}. In geometric optics, their combined magnification diverges as $\mu\propto\delta^{-1/2}$, where $\delta$ is the perpendicular source--caustic distance. In Fig.~\ref{fig:map}, the fold corresponds to a local maximum of the source azimuth $\phi_s$ as a function of the image parameter $\varphi$: the two solutions for $\varphi$ at fixed $\phi_s$ coalesce, rather than two source-azimuth values.
This point-source, geometric-optics divergence is regularized by diffraction or, under suitable assumptions, finite-source averaging. Here, $\mu$ denotes intensity or squared-strain magnification.

\paragraph*{Finite source size.}
For spatially incoherent emission, the extended-source magnification is the surface-brightness-weighted average of the point-source response. Near an isolated fold, a sufficiently small source with effective transverse radius $R_*$ and a fixed brightness profile has
\bea\label{eq:finitemu}
\mu_{\rm max}^{\rm (finite)}
\sim C_{\rm f}\left(\frac{R_*}{R_g}\right)^{-1/2}\!\!,
\eea
where $C_{\rm f}$ depends on the local fold strength and source profile. The peak need not coincide with the source center crossing the caustic.
For a b-EMRI, however, the radiation is coherent, requiring an amplitude-level treatment that retains the source's spatial phase structure. The binary separation $d$ cannot therefore be identified directly with $R_*$. The estimate $\mu_{\rm max}^{\rm (finite)}\propto(d/R_g)^{-1/2}$ applies only if the source model justifies both $R_*\sim d$ and the averaging prescription.

\paragraph*{Wave-optics effects.}
Near a generic fold, the cubic phase dependence along the degenerate direction yields an Airy-type diffraction integral \cite{Nakamura:1997sw,petters2001singularity}. In the high-frequency fold approximation, the peak magnification scales as \cite{Takahashi:2003ix,Matsunaga:2006uc}
\bea\label{eq:wavemu}
\mu_{\rm max}^{\rm (wave)}
\propto \left(\frac{\lambda_{\rm GW}}{R_g}\right)^{-1/3}\!\!,
\eea
at fixed dimensionless lens geometry. The coefficient depends on the local lens mapping. Diffraction thus replaces the geometric-optics divergence with a finite Airy pattern.

These effects provide complementary constraints on the source and lens. When finite-source averaging is valid, Eq.~\eqref{eq:finitemu} probes the effective source size together with the fold strength; the frequency-dependent diffraction pattern instead probes combinations of the lens scale and local mapping coefficients. Neither scaling alone determines $d$ or $R_g$, and their exponents describe different physical dependences rather than a direct observational discriminator.

The estimates in Sec.~\ref{sec:oom} suggest that the source-plane diffraction width exceeds the projected binary separation for representative b-EMRI parameters. If the effective coherent emitting region is also smaller than this width, the point-source Airy response provides a useful leading approximation. A treatment incorporating both diffraction and the spatial amplitude and phase structure of the binary's radiation is left for future study.

\begin{figure*}[htbp]
\centering
\hspace*{-5mm}
    \includegraphics[width=2\columnwidth]{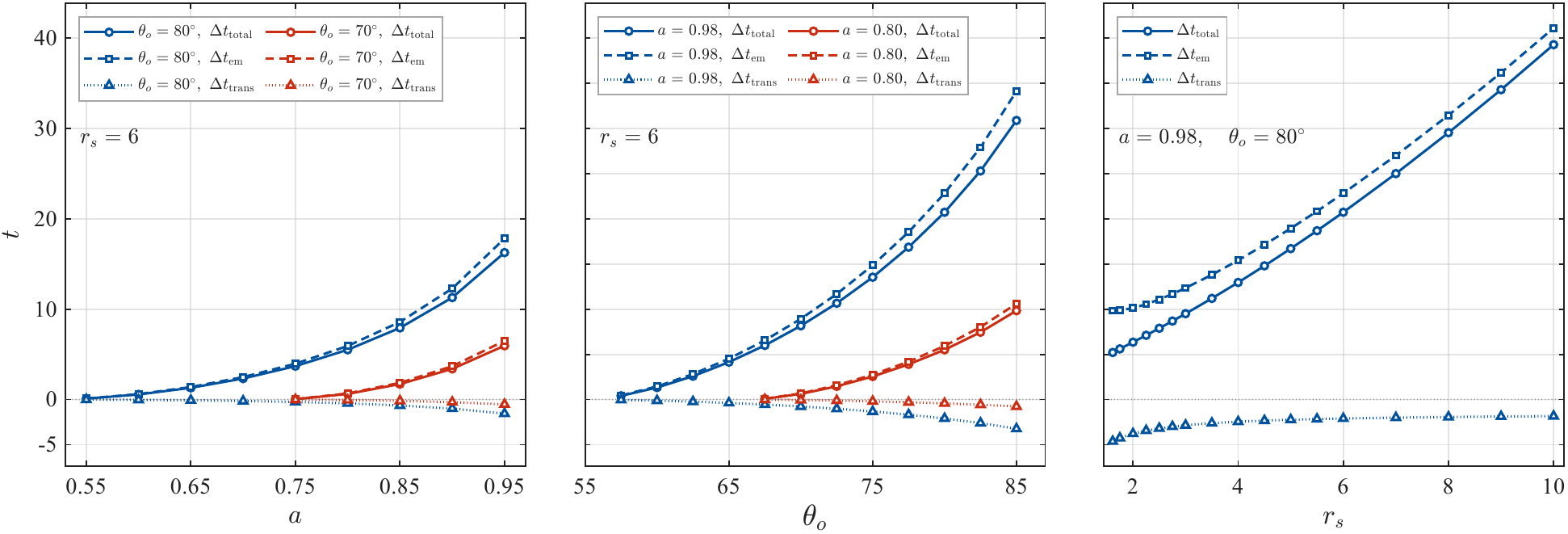}
\caption{Time differences between the paired flashes for different values of $a$, $\t_o$, and $r_s$, for the prograde case. The total time difference $\D t_{\rm tot}$ is shown alongside the contributions from source motion, $\D t_{\rm em}$, and lensed photon propagation, $\D t_{\rm trans}$. }
    \label{fig:Dtime}
\end{figure*}

\section{Magnifications and flashes}\label{sec:magnif}

The magnification factor is jointly determined by the backfround spacetime, the observer's inclination, and the source's motion, as predicted by Eq.~\eqref{eq:GeometricMagnification}. 
Here we model the inner binary as a point source on a Keplerian orbit around the central black hole. Its four-velocity is
$u_{\mu}=\left(-E_K,\,0,\,0,\,L_K\right)$,
where $E_K$ and $L_K$ are the specific orbital energy and angular momentum \cite{Bardeen:1972fi}. Ray tracing and the associated Jacobi map determine the geometric magnification of each path, with source motion entering through the source-frame normalization and angular mapping.

Fig.~\ref{fig:mumap} shows the magnification in Bardeen screen coordinates (top panels, $\ln\mu$) and as a function of screen polar angle (bottom panels, $\mu$). Along the source's outer orbit, the direct image exhibits a pronounced approaching--receding asymmetry. For the prograde orbit shown in the top panels, $\mu$ exceeds $2$ on the approaching side (left) and falls to approximately $0.5$ on the receding side (right). This asymmetry reflects the mapping from source-frame emission directions to the observer's screen, which depends on both source motion and curved-spacetime propagation. Strong-field focusing further modifies the magnification near the black hole. Gravitational and kinematic frequency shifts affect the observed energy flux~\cite{Zhang:2023cdh}, but are distinct from ray-bundle magnification.

The lensed image ($N=1$) is generally fainter than the direct image, but its magnification is set by the local lens map rather than by the solid angle subtended by its rays in the source sky~\cite{Bozza:2002zj}. In the strong-deflection limit, successive higher-order images approaching the same critical-curve segment are approximately exponentially demagnified:
\bea
\mu_{N+1} \,\sim\, e^{-\gamma(\tilde{r})}\,\mu_N\,,
\eea
where $\gamma(\tilde{r})$ is the Lyapunov exponent governing the demagnification associated with the unstable spherical photon orbit at $r=\tilde{r}$. This asymptotic scaling need not apply to the lowest-order images or near caustics. Where it applies, the corresponding strain-amplitude ratio is approximately $e^{-\gamma(\tilde{r})/2}$.

The bottom panels of Fig.~\ref{fig:mumap} show $\mu$ as a function of screen polar angle $\varphi$ at fixed source radius $r_s$. Away from caustics, the angular magnification profiles of the direct image and the lower-screen portion of the lensed image change little as the spin increases from $a=0.8$ to $0.98$; their broad approaching--receding asymmetries remain largely unchanged. At high observer inclination, the lower-screen portion of the lensed image exhibits a smooth angular trend resembling that of the direct image, although the two branches have distinct ray trajectories and magnifications.

By contrast, the upper-screen portion of the lensed image exhibits a more intricate pattern, with narrow caustic peaks that shift and change in amplitude with spin. Its magnification depends sensitively on the mapping between source-frame emission directions and screen positions, which is shaped by strong deflection and frame dragging. The observed asymmetry therefore cannot be attributed to screen position alone.

\begin{table*}[t]
\caption{\label{table:ttotal}
\protect\parbox[t]{0.96\textwidth}{\raggedright
Estimated time separations of the paired flashes 
using Bozza's formula and large-$r_s$ approximation (Eq.~\eqref{eq:timed}). The black-hole mass is $M=10^{6}M_{\odot}$, and $r_s=2\,\mathrm{ISCO}_{\mathrm{pro}}(a)$ for each spin.}}
\vspace{0.7ex}
\begingroup
\centering
\small
\setlength{\tabcolsep}{3pt}
\setlength{\fboxsep}{2pt}
\setlength{\fboxrule}{0.35pt}
\renewcommand{\arraystretch}{1.32}
\newcommand{\timecell}[1]{%
  \fbox{\makebox[4.2em][c]{$#1$}}%
}
\begin{tabular}{@{}cc@{\hspace{9pt}}cccc@{\hspace{12pt}}cccc@{}}
\toprule
& &
\multicolumn{4}{c}{$m=2$: \,$\Delta t_{\mathrm{total}}$ (second)} &
\multicolumn{4}{c}{$m=3$: \,$\Delta t_{\mathrm{total}}$ (second)} \\
\cmidrule(lr){3-6}\cmidrule(l){7-10}
$a/M$ & $\mathrm{ISCO}_{\mathrm{pro}}/M$
& $60^\circ$ & $70^\circ$ & $80^\circ$ & $90^\circ$
& $60^\circ$ & $70^\circ$ & $80^\circ$ & $90^\circ$ \\
\midrule
0.6   & 3.829
& \timecell{\varnothing} & \timecell{\varnothing}
& \timecell{0.06}       & \timecell{38.70}
& \timecell{\varnothing} & \timecell{4.09}
& \timecell{35.30}      & \timecell{104.30} \\
0.7   & 3.393
& \timecell{\varnothing} & \timecell{\varnothing}
& \timecell{3.77}       & \timecell{43.90}
& \timecell{\varnothing} & \timecell{16.80}
& \timecell{52.20}      & \timecell{118.40} \\
0.8   & 2.907
& \timecell{\varnothing} & \timecell{\varnothing}
& \timecell{8.40}       & \timecell{45.50}
& \timecell{2.91}       & \timecell{29.30}
& \timecell{63.70}      & \timecell{122.70} \\
0.9   & 2.321
& \timecell{\varnothing} & \timecell{0.03}
& \timecell{11.20}      & \timecell{41.10}
& \timecell{10.10}      & \timecell{35.40}
& \timecell{64.40}      & \timecell{110.80} \\
0.998 & 1.237
& \timecell{\varnothing} & \timecell{0.92}
& \timecell{6.83}       & \timecell{19.70}
& \timecell{8.42}       & \timecell{20.50}
& \timecell{33.30}      & \timecell{53.00} \\
\bottomrule
\end{tabular}
\par
\endgroup
\vspace{0.7ex}
\noindent
\begin{minipage}{\textwidth}
\raggedright\footnotesize
\textit{Note.} Column angles are observer inclinations $\theta_o$.
$\varnothing$ denotes failure of the real-valued applicability
condition of Eq.~\eqref{eq:thetac}. The listed
$\Delta t_{\mathrm{total}}$ values use the approximation
$\Delta t_{\mathrm{total}}\simeq\Delta t_{\mathrm{em}}$ and exclude
the propagation-time contribution.
For $M=10^{6}M_{\odot}$, $1M\simeq4.92549\,\mathrm{s}$.
\end{minipage}
\end{table*}

Each caustic crossing can produce a transient amplification peak if the source is emitting at that time. In Fig.~\ref{fig:mumap}, for example, the sharp spikes in the $N=1$ curve near $\varphi\approx50^\circ$ and $150^\circ$ correspond to the source crossing folds of the CS associated with the observer's past light cone. When emission persists across both crossings, the resulting peaks are separated in arrival time by the interval between emission events plus the difference in their propagation times (Fig.~\ref{fig:Dtime}):
\bea\label{eq:timediff}
\begin{aligned}
&\Delta t_{\rm total} = \Delta t_{\rm em} + \Delta t_{\rm trans} \,, \\
&\Delta t_{\rm em} = \f{\phi_s^{(b)}-\phi_s^{(a)}}{\Omega_K},\quad
\Delta t_{\rm trans} = \Delta t_{(b)}-\Delta t_{(a)}\,,
\end{aligned}
\eea
where $(a)$ and $(b)$ label successive crossing events, $\Omega_K=(r_s^{3/2}\pm a)^{-1}$ is the Keplerian angular velocity (with $+$ for prograde and $-$ for retrograde motion), and $\Delta t_{(a,b)}$ denote the corresponding propagation times to the observer. The azimuthal difference $\phi_s^{(b)}-\phi_s^{(a)}$ is measured along the direction of orbital motion.

For a specified orbital orientation and image branch, Eq.~\eqref{eq:timediff} relates the peak separation to the spin $a$, observer inclination $\t_o$, and source radius $r_s$. Fig.~\ref{fig:Dtime} displays the two contributions separately for prograde orbits. Within the plotted parameter ranges where two crossings exist, the total separation increases with $a$, $\t_o$, and $r_s$. The emission-time contribution dominates these examples, while the propagation contribution becomes more important at small source radii for rapidly rotating black holes viewed at high inclination. In particular, changes in spin and inclination alter the azimuthal positions of the equatorial caustic intersections; their effects on the peak separation cannot be inferred solely from the size of the critical curve on the observer's sky.

The emission-time contribution is governed by the azimuthal
separation of the caustic crossings, which approaches a
constant at large source radius
(Fig.~\ref{fig:3DCS}).
Using Bozza's small-spin approximation
\cite{Bozza:2008mi}, we obtain
\bea\label{eq:dphiinf}
\D\phi_s^{(\infty)} \,\approx\, 2\,|C_m|\sin{\t_o}
\left[1-\left|\frac{\cos{\t_o}}
{C_m\sin^3{\t_o}}\right|^{2/3}\right]^{3/2},
\eea
where
$C_m=a^2(5m\pi+8\sqrt{3}-36)/18$
in the large-radius limit, with finite-radius corrections
of order $r_s^{-1}$.
This expression applies at fixed caustic order $m$
and requires a nonnegative square-bracketed factor.
In the edge-on limit, $\t_o\to90^\circ$,
$\D\phi_s^{(\infty)}\to2|C_m|\propto a^2$.
For the secondary caustic ($m=2$),
$2|C_2|\simeq1.03\,a^2$, so that
$\D\phi_s^{(\infty)}\approx a^2$.

For a source on a circular equatorial orbit, substituting
the asymptotic angular separation into
Eq.~\eqref{eq:timediff} yields
\bea\label{eq:timed}
\Delta t_{\rm em}\,\approx\,
\D\phi_s^{(\infty)}\,r_s^{3/2},
\qquad r_s\gg1,
\eea
with radius, spin, and time expressed in units of the
SMBH mass.
Thus, for the secondary caustic and a sufficiently
edge-on observer,
$\Delta t_{\rm em}\approx a^2r_s^{3/2}$.
Here, ``sufficiently edge-on'' requires
$|\cos{\t_o}|\ll|C_2|\sin^3{\t_o}$; proximity to
$90^\circ$ alone is not sufficient uniformly as
$a\to0$.
Away from this limit, the square-bracketed factor in
Eq.~\eqref{eq:dphiinf} introduces additional spin and
inclination dependence and can eliminate distinct
source--caustic crossings.

An equatorial intersection of the caustic surface does
not by itself imply paired crossings of a given source
orbit.
For the configuration considered here,
Eq.~\eqref{eq:Ncomp} requires $a>a_c(\t_o)$ for crossings
on two distinct caustic branches.
In the large-radius approximation, the
boundary occurs when the square-bracketed factor in
Eq.~\eqref{eq:dphiinf} vanishes; distinct crossings
require it to be positive.
\footnote{We exclude here the regime $a>a_0(\t_o)$
in Eq.~\eqref{eq:Ncomp}, in which the secondary caustic
surface intersects the equatorial plane at finite
radius. The time separation is also small in the
configurations considered in that regime.}

For the secondary caustic, the asymptotic boundary
satisfies $|C_2|\sin^3{\t_{\rm c}}=|\cos{\t_{\rm c}}|$.
Using $|C_2|\approx a^2/2$ and restricting to
$0\leq\t_{\rm c}\leq90^\circ$ gives
\bea\label{eq:thetac}
a^2\sin^3{\t_{\rm c}}\approx2\cos{\t_{\rm c}}.
\eea
This relation estimates the large-radius boundary,
rather than the finite-radius threshold
$\t_{\rm c}(a,r_s)$, and is valid only within the
small-spin approximation.
For the $m=2$, $r_s=6$ configurations in the middle
panel of Fig.~\ref{fig:Dtime}, the numerical thresholds
are $\t_{\rm c}\approx57^\circ$ at $a=0.98$ and
$\t_{\rm c}\approx67^\circ$ at $a=0.8$.
Although the growth of $|C_m|$ with $m$ formally suggests
$\t_{\rm c}\to0^\circ$, the fixed-order small-spin
expansion does not support extrapolation to arbitrarily
high caustic orders.

The propagation contribution $\Delta t_{\rm trans}$
requires separate consideration.
As illustrated in Fig.~\ref{fig:CSimage}, the images
associated with the two crossings lie near distinct
points on the critical curve, corresponding to unstable
spherical photon orbits at $\tilde r_{(a)}$ and
$\tilde r_{(b)}$.
The characteristic time increments $\tilde\tau_{(a)}$
and $\tilde\tau_{(b)}$ quantify coordinate-time
accumulation near these orbits \cite{Gralla:2019drh}
and provide relevant scales for the strong-field delay.
The difference between the complete propagation times,
however, also depends on how long each ray remains near
its associated orbit and on the rest of its trajectory.
Thus, neither
$\Delta t_{\rm trans}\propto\tilde\tau_{(b)}-\tilde\tau_{(a)}$
nor a finite large-$r_s$ limit follows from the
critical-orbit correspondence alone; either conclusion
would require an explicit asymptotic calculation.

Table~\ref{table:ttotal} gives estimates for an SMBH
with $M=10^6M_{\odot}$ at several spins and observer
inclinations.
Because these estimates use the large-radius,
small-spin approximation and neglect
$\Delta t_{\rm trans}$, they should be interpreted as
rough estimates rather than precise predictions of the
total arrival-time separation.
For both flashes to be emitted before the inner binary
coalesces, its remaining lifetime at the first crossing
must exceed the emission-time interval
$\Delta t_{\rm em}$, with both times expressed in the
same time coordinate.
Using the leading-order coalescence time for an
equal-mass circular binary and taking
$r_s\sim\mathcal{O}(10M)$ gives the order-of-magnitude
condition
\bea
d^4 M^3 \gtrsim 10^{7/2} \,m^3 J^2 \sin{\t_o} \,,
\eea
where $J$ is the SMBH angular momentum and we have
used the nearly edge-on approximation.
A remaining lifetime comparable to the outer orbital
period would favor a single paired-flash episode;
a longer-lived source could instead produce recurring
pairs over successive orbits.

A peak separation alone does not establish that the signature is detectable. Detectability also depends on the source waveform and its duration in the detector band, the relative amplitudes and signal-to-noise ratios of the relevant images, and the statistical criteria used to associate potential lensing pairs \cite{Li:2018prc, Hannuksela:2019kle, LIGOScientific:2021izm, Caliskan:2022hbu, Chan:2024qmb}. If the images overlap in time, lensing may instead be inferred from specific distortions of the combined waveform; if they are temporally separated, both images must be individually detectable, and their recovered waveforms must be consistent with a common physical source. A time delay that is not excessively long may facilitate such an association, but it does not, by itself, imply robust detectability with current or forthcoming ground-based detectors.

Finally, the near-horizon regime may offer further analytical insight into caustic crossings, including those associated with emission during plunge, which we do not study here. In the extremal limit, near-horizon scaling yields the NHEK geometry \cite{Bardeen:1999px}, whose enhanced symmetry simplifies particle and field dynamics \cite{Lupsasca:2014pfa,Gralla:2017ufe,Kapec:2019hro,Hou:2023hto}. It may likewise facilitate an analytical treatment of the Jacobi map and amplification peaks, though predictions for distant observers would require matching to the exterior Kerr geometry.

\section{Summary and discussion}\label{sec:discussion}

We have developed a geometric-optics framework for GWs emitted by a localized source orbiting a Kerr black hole, motivated by self-lensing in b-EMRIs. Point-to-point null geodesics determine image branches and propagation times, while bilocal geodesic operators yield the Jacobi map governing ray-bundle focusing. Together, these constructions relate image multiplicity and Kerr caustic geometry to image-dependent amplification. For regular images, matched directional emission normalizations give a geometric strain-envelope enhancement of $\sqrt{\mu}$, with source evolution and image-dependent frequency shifts treated separately. Although applied here to circular equatorial orbits, the framework also accommodates other localized emitters in the geometric-optics regime.

The extended Kerr caustics couple transient amplification to orbital motion (Fig.~\ref{fig:3DCS}). Depending on the black-hole spin, observer inclination, and orbital radius, a circular equatorial orbit can cross two folds of the secondary caustic at distinct azimuths. If emission persists through both crossings, the source can produce paired flashes within one revolution (Fig.~\ref{fig:mumap}); sustained emission over multiple orbits can produce recurring pairs. These episodes enhance higher-order images that are typically demagnified away from caustics.

The arrival-time separation is the sum of the emission-time interval between crossings and the difference in propagation times. The emission-time interval dominates most configurations shown in Fig.~\ref{fig:Dtime}, whereas propagation delays become more important closer to the black hole. Table~\ref{table:ttotal} estimates the emission-time contribution from a large-radius caustic approximation; the total separation additionally requires the differential propagation delay. The central black-hole mass sets the characteristic lensing length and time scales, while the inner binary's properties and evolution determine its GW frequency and emission duration.

Paired flashes provide a potential timing and waveform signature of orbital caustic crossings. Because they originate from distinct emission events, associating them requires a model of the source evolution; consistency with the predicted phase evolution can provide an additional test \cite{Ezquiaga:2023xfe}. Image contributions that overlap at the detector must be summed coherently at the strain level. If multiple branches are measurable, their emission directions, magnifications, frequency shifts, delays, and polarization transport may provide complementary constraints on the source and Kerr geometry. Their information content must therefore be assessed through a joint waveform analysis that accounts for detector response, signal duration, and noise, rather than by treating individual branches as independent detectors.

The present calculation supplies the ray-based propagation component of such an analysis. Geometric optics breaks down at caustics, where finite amplification profiles require a uniform wave-optical treatment of diffraction and interference \cite{Caliskan:2022hbu,Tambalo:2022plm,Leung:2023lmq,Pijnenburg:2024btj,Ezquiaga:2025gkd}. For an inner binary, this treatment must preserve the coherent amplitude and phase structure of the emitted radiation; the binary separation cannot simply be used as an incoherent source size to smooth the geometric-optics divergence. Beyond leading-order geometric optics, the gravitational spin Hall effect can also produce frequency- and helicity-dependent corrections to propagation \cite{Oancea:2022szu, Oancea:2023hgu}. These extensions would relate the ray-based caustic-crossing pattern to finite-frequency flash profiles.

A natural next step is a waveform calculation combining evolving binary emission, Kerr propagation, wave optics near caustics, and detector response. Such a calculation would establish the detectability of paired flashes and quantify their contribution to parameter inference. The central result here is the connection between orbital crossings of Kerr caustics and paired, potentially recurring amplification episodes. With consistent waveform modeling, this pattern may provide complementary constraints on the source orbit and the central black-hole geometry.

\medskip

\section*{acknowledgements}
We thank X. Chen for the helpful discussion.
The work is partly supported by NSFC Grant No.12275004, 12205013, 12575048, 12547123, 12588101. M. Guo is also supported by the BNU Tang Scholar. 
	
\textbf{Statement on the use of AI.\,}
The ideas, modeling, calculations and manuscript were developed by the human authors. OpenAI's ChatGPT 5.6 assisted with language editing, figure refinement, and numerical consistency checks. The authors reviewed all AI-assisted revisions and take full responsibility for the manuscript.

\medskip

\appendix

\section{Jacobi maps and physical magnification}
\label{app:JacobiFlux}

The strain-envelope magnification introduced in the main text follows
from the transverse area of a null-geodesic bundle. Here we derive this
area from the bilocal geodesic operators, relate it to the radiation of
a point source, and evaluate the resulting magnification using the Kerr
lens map. We use the curvature
convention
\begin{equation}
 [\nabla_\alpha,\nabla_\beta]v^\mu
 =-R^\mu{}_{\nu\alpha\beta}v^\nu.
 \label{eq:appCurvatureConvention}
\end{equation}
The propagation is treated in lossless geometric optics. Unless stated
otherwise, the amplitude and flux refer to one image branch away from
a caustic.

\subsection{Bilocal geodesic operators}
\label{app:JacobiFoundation}
Consider a fiducial null geodesic with future-directed affine tangent
$k^\mu=\md x^\mu/\md\lambda$, connecting the source at $\lambda_s$ to
the observer at $\lambda_o$. For a smooth family of neighboring
geodesics $x^\mu(\lambda,\varepsilon)$, define the connecting vector
and its covariant derivative by
\begin{equation}
 \begin{aligned}
 \xi^\mu=\left.\frac{\partial x^\mu}{\partial\varepsilon}
                  \right|_{\varepsilon=0},\quad
 \ell^\mu=\mathrm{D}_\lambda\xi^\mu
 =\delta k^\mu+\Gamma^\mu{}_{\alpha\beta}k^\alpha\xi^\beta.
 \end{aligned}
 \label{eq:appCovariantVariation}
\end{equation}
with $ \mathrm{D}_\lambda=k^\alpha\nabla_\alpha$ denotes the covariant derivative. The connection term is needed when the initial event varies. Since $[k,\xi]=0$, the first variation of the geodesic equation gives
\begin{equation}
 \begin{aligned}
 \mathrm{D}_\lambda\xi^\mu=\ell^\mu,\quad
 \mathrm{D}_\lambda\ell^\mu=\mR^\mu{}_{\nu}\xi^\nu.
 \end{aligned}
 \label{eq:appFirstVariation}
\end{equation}
where $
 \mR^\mu{}_{\nu}=R^\mu{}_{\alpha\nu\beta}k^\alpha k^\beta$ represents the tidal tensor.
This expansion is first order in the separation of the rays, it does not require the accumulated curvature along the fiducial geodesic to be small \cite{Grasso:2018mei,Serbenta:2021tzv}.

The bilocal geodesic operators (BGOs) propagate the initial position and tangent deviations according to
\begin{equation}
 \begin{pmatrix}\xi^\mu(\lambda)\\ \ell^\mu(\lambda)\end{pmatrix}
 =
 \begin{pmatrix}
 W_{XX}^{\mu}{}_{\nu'}&W_{XK}^{\mu}{}_{\nu'}\\
 W_{KX}^{\mu}{}_{\nu'}&W_{KK}^{\mu}{}_{\nu'}
 \end{pmatrix}
 \begin{pmatrix}\xi^{\nu'}(\lambda_s)\\
                 \ell^{\nu'}(\lambda_s)\end{pmatrix}.
 \label{eq:appBGO8}
\end{equation}
Primed indices refer to the initial event. Each block is a map between four-dimensional tangent spaces, and the full propagator $W$ acts on eight-dimensional position--tangent data. Taking the block components of the evolution equation for \(W\) gives
\be
\begin{aligned}
&\mathrm{D}_{\lambda}W_{XX}^{\mu}{}_{\nu'}
    =W_{KX}^{\mu}{}_{\nu'},\quad\quad\,\,
\mathrm{D}_{\lambda}W_{XK}^{\mu}{}_{\nu'}
    =W_{KK}^{\mu}{}_{\nu'},\\
&\mathrm{D}_{\lambda}W_{KX}^{\mu}{}_{\nu'}
    =\mathcal{R}^{\mu}{}_{\rho}
      W_{XX}^{\rho}{}_{\nu'},\,\,\,
\mathrm{D}_{\lambda}W_{KK}^{\mu}{}_{\nu'}
    =\mathcal{R}^{\mu}{}_{\rho}
      W_{XK}^{\rho}{}_{\nu'}.
\end{aligned}
\label{eq:appBGOBlockEvolution}
\ee
Here \(\mathrm{D}_{\lambda}\) acts on the unprimed index, while the primed
index labels the initial tangent space and is held fixed. Since \(W\)
propagates the initial position--tangent data unchanged at
\(\lambda=\lambda_s\), its initial conditions are
\begin{equation}
\begin{aligned}
W_{XX}^{\mu}{}_{\nu'}(\lambda_s)
    &=\delta^{\mu}{}_{\nu'},&
\quad
W_{XK}^{\mu}{}_{\nu'}(\lambda_s)&=0,\\
W_{KX}^{\mu}{}_{\nu'}(\lambda_s)&=0,&
\quad
W_{KK}^{\mu}{}_{\nu'}(\lambda_s)
    &=\delta^{\mu}{}_{\nu'}.
\end{aligned}
\label{eq:appBGOBlockInitialConditions}
\end{equation}
The full propagator is
invertible and composes along successive geodesic segments. We use
its transverse reduction below; the general bilocal construction is
discussed in Refs.~\cite{Grasso:2018mei,Serbenta:2021tzv}.

\subsection{Jacobi matrix and optical scalars}
\label{app:OpticalDeformation}
Choose an orthonormal source screen basis $e_A{}^\mu$, with $A=1,2$, and parallel transport it along the ray.
Projection of $W$ onto the transported basis gives the reduced propagator
\begin{equation}
 \begin{aligned}
 W_\perp=\begin{pmatrix}\mathcal C&\mD\\\mathcal C'&\mD'\end{pmatrix},\quad
 W_\perp'=\begin{pmatrix}0&\mathbb I_2\\\mR&0\end{pmatrix}W_\perp,
 \end{aligned}
 \label{eq:appBGO4}
\end{equation}
where the screen tidal matrix is
\begin{equation}
 \mR_{AB}=R_{\mu\alpha\nu\beta}
 e_A{}^\mu k^\alpha e_B{}^\nu k^\beta=\mR_{BA}.
 \label{eq:appScreenTidal}
\end{equation}
The $4\times4$ matrix $W_\perp$ acts on transverse phase space, it is not one of the $4\times4$ spacetime blocks of $W$. Its upper-right block is the Jacobi matrix used in the main text,
\begin{equation}
 \mD^A{}_{B}(\lambda,\lambda_s)
 =e^A{}_{\mu}(\lambda)W_{XK}^{\mu}{}_{\nu'}(\lambda,\lambda_s)
 e_B{}^{\nu'}(\lambda_s).
 \label{eq:appJacobiProjection}
\end{equation}
It satisfies
\begin{equation}
 \mD''=\mR\mD,
 \label{eq:appJacobiInitial}
\end{equation} 
where the initial condition is $\mD(\lambda_s,\lambda_s)=0, \mD'(\lambda_s,\lambda_s)=\mathbb I_2$.
At the source, the complementary block satisfies $\mathcal C=\mathbb I_2$
and $\mathcal C'=0$. The source-vertex condition $\xi^\mu(\lambda_s)=0$
then leaves $\mD$ as the only block contributing to the transverse
separation.

The longitudinal solutions of the Jacobi equation imply
\be
 \begin{aligned}
 & k_\mu(\lambda)W_{XK}^{\mu}{}_{\nu'}(\lambda,\lambda_s)
=(\lambda-\lambda_s)k_{\nu'}(\lambda_s),\\
 & W_{XK}^{\mu}{}_{\nu'}(\lambda,\lambda_s)k^{\nu'}(\lambda_s)
=(\lambda-\lambda_s)k^\mu(\lambda).
 \end{aligned}
 \label{eq:appNullContractions}
\ee
Consequently, Eq.~\eqref{eq:appJacobiProjection} is well defined on the
transverse quotient. Rotating the endpoint screen bases or changing
their observer representatives leaves $|\det\mD|$ unchanged at fixed
ray, endpoints, and affine normalization. A change of intersection
hypersurface similarly adds $k^\mu\delta\lambda$ to the endpoint
variation, without changing its screen projection.

On a regular segment, the optical deformation matrix is
\begin{equation}
 \mathcal B=\mD'\mD^{-1},\quad
 \label{eq:appDeformation}
\end{equation}
It acts on the current separation, $\xi'=\mathcal B\xi$.
Symmetry of the screen tidal matrix implies
\begin{align}\label{eq:appWronskian}
\frac{\md}{\md\lambda}
\left(\mD^T\mD'-\mD'^T\mD\right)=0.
\end{align}
The vertex initial conditions set this conserved
matrix to zero, hence
$\mathcal{B}=\mathcal{B}^T$ wherever $\mD$ is
invertible. A vertex bundle is therefore twist
free on each regular segment. Defining
$\theta=\operatorname{tr}\mathcal{B}/2$ and
$\sigma_{AB}=\mathcal{B}_{AB}-\theta\,\delta_{AB}$
gives
\begin{align}
\label{eq:appOpticalEvolution}
\begin{aligned}
&\mathrm{D}_{\lambda}\mathcal{D}
=2\,\theta\,,\\
&\mathcal{B}'+\mathcal{B}^2=\mathcal{R}\,,\\
&\theta'
=
-\theta^2-\frac12\sigma_{AB}\sigma^{AB}
+\frac12\operatorname{tr}\mathcal{R}\,,
\end{aligned}
\end{align}

In a vacuum black-hole exterior,
$\operatorname{tr}\mathcal{R}=0$, although
$\mathcal{R}$ generally remains nonzero. The Weyl
tidal field generates shear, which modifies the
area expansion through the quadratic term in
Eq.~(\ref{eq:appOpticalEvolution}). Vacuum focusing
therefore reflects the accumulated action of a
trace-free tidal field, not a nonzero local tidal
trace.

Magnification compares accumulated area mappings,
not instantaneous expansion: a magnified bundle
may still be expanding at reception. The singular
values of $\omega_s\mD$ give the principal transverse
stretches of a small source angular patch, whereas
their product gives its area factor. Equal
determinants need not imply equal beam shapes.

\subsection{Point-source radiation}
\label{app:FluxTransport}
\label{app:PhysicalArea}

An angular displacement $\md\vartheta^B$ in the source rest frame,
at fixed $\omega_s$, gives
$\delta k_s^\mu=\omega_s e_B{}^\mu\md\vartheta^B$. The Jacobi matrix
therefore determines the receiving displacement and transverse area:
\begin{equation}
\xi_o^A=\omega_s\mD_s^A{}_{B}\md\vartheta^B,\quad \md\mathscr A_o=\omega_s^2|\det\mD_s|\,\md\Omega_s.
 \label{eq:appArea}
\end{equation}
The angular-to-area map of Eq.~\eqref{eq:appArea}
determines the source-solid-angle area distance $D_{\rm G}$. The observer-vertex map similarly gives the angular-diameter distance
$D_{\rm A}$. With reciprocity,
\begin{equation}
 \begin{aligned}
 D_{\rm G}^2&=\frac{\md\mathscr A_o}{\md\Omega_s}
             =\omega_s^2|\det\mD_s|,\\
 D_{\rm A}^2&=\frac{\md\mathscr A_s}{\md\Omega_o^{\rm sky}}
             =\omega_o^2|\det\mD_o|.
 \end{aligned}
 \label{eq:appAreaDistances}
\end{equation}
Obviously, the two satisfy the relation $ D_{\rm G}^2=D_{\rm A}^2/g^2$. Changing the source rest frame changes its angular measure by $\md\widetilde\Omega_s=(\omega_s/\widetilde\omega_s)^2\md\Omega_s$, while leaving the receiving area of the same bundle unchanged.
Source aberration is thus already included in the angular
normalization of $D_{\rm G}$ \cite{Grasso:2018mei}.

For a point source, define the directional source-frame spectral
luminosity by
\begin{equation}
 \mathcal L_{\nu_s}(\tau_s,n_s)
 =\frac{\md E_s}{\md\tau_s\,\md\nu_s\,\md\Omega_s}.
 \label{eq:appDirectionalLuminosity}
\end{equation}
Here $\tau_s$ is source proper time and $n_s$ is the emission direction
in its rest frame. The emission can be anisotropic and time dependent.
For an isotropic source, $\mathcal L_{\nu_s}=L_{\nu_s}/(4\pi)$,
where $L_{\nu_s}$ is the luminosity per unit frequency, integrated
over emission directions.

For an infinitesimal packet transported without loss along one image
branch, the energy, frequency interval, and proper-time interval satisfy
\begin{equation}
 \md E_o=g\,\md E_s,\quad
 \md\nu_o=g\,\md\nu_s,\quad
 \md\tau_o=\md\tau_s/g.
 \label{eq:appPacketTransform}
\end{equation}
The spectral flux is the received energy per unit proper time,
frequency, and transverse area,
$F_{\nu_o}=\md E_o/(\md\tau_o\,\md\nu_o\,\md\mathscr A_o)$.
Together with Eq.~\eqref{eq:appAreaDistances}, this gives
\begin{align}
 F_{\nu_o}(\tau_o)
 &=\frac{g}{\omega_s^2|\det\mD_s|}
 \left.\mathcal L_{\nu_s}\bigl(\tau_s(\tau_o),n_s\bigr)
 \right|_{\nu_s=\nu_o/g}\nonumber\\
&=\frac{g}{D_{\rm G}^2}\mathcal L_{\nu_s}
    =\frac{g^3 }{D_{\rm A}^2}\mathcal L_{\nu_s}.
 \label{eq:appIntegratedFlux}
\end{align}

\subsection{Kerr lens map and point-source flux}
\label{app:KerrLensFlux}
\label{app:KerrIntegralMap}

We use the Kerr potentials, path integrals, and Bardeen screen coordinates of Sec.~\ref{sec:KerrLensing}.
The source follows a circular equatorial geodesic at $r_s$. Its
four-velocity $u_{s\mu}=(-E_K,0,0,L_K)$ can be written as
\begin{equation}
 \begin{aligned}
& u_s^\mu=u_s^t[(\partial_t)^\mu+\Omega_K(\partial_\phi)^\mu],\\
& \Omega_K=\frac{\pm1}{r_s^{3/2}\pm a},\quad
 u_s^t=\frac{r_s^{3/2}\pm a}
 {\sqrt{r_s^3-3r_s^2\pm2a r_s^{3/2}}}.
 \end{aligned}
 \label{eq:appCircularOrbit}
\end{equation}
For $0\leq a<1$, the upper and lower signs denote prograde and
retrograde motion, respectively \cite{Bardeen:1972fi}. The orbit is outside the horizon and timelike. For a static observer at infinity,
\begin{equation}
 g=\frac{1}{u_s^t(1-\Omega_K l)}.
 \label{eq:appCircularRedshift}
\end{equation}
Denote the radial and polar integrals in Eq.~\eqref{eq:lenseqs} by
\begin{equation}
 I_r=\fint_{r_s}^{r_o}\frac{\md r}{v_r\sqrt R},\quad
 G_\theta=\fint_{\theta_s}^{\theta_o}
                      \frac{\md\theta}{v_\theta\sqrt\Theta},
 \label{eq:appKerrPathIntegrals}
\end{equation}
and define
\begin{equation}
 A=I_r-G_\theta,\quad B=\Delta\phi-\Omega_K\Delta t.
 \label{eq:appABFunctions}
\end{equation}
The quantities $\Delta\phi$ and $\Delta t$ retain the radial--polar
splitting of Eq.~\eqref{eq:lenseqs}. The slash includes all monotonic
segments, with the signs changing at the corresponding turning
points. As in the main text, $N$, $m$, and $w$ count equatorial crossings,
polar turning points, and radial turning points, respectively. A local image branch fixes the turning history and
signs, with the azimuth unwrapped continuously. The integrals admit
elliptic representations on each such branch
\cite{Gralla:2019ceu}.

At a fixed observation event $O$, introduce the corotating azimuths
\begin{equation}
 \psi_s=\phi_s-\Omega_Kt_s,\quad
 \psi_o=\phi_o-\Omega_Kt_o.
 \label{eq:appCorotatingAzimuth}
\end{equation}
The lens equations are then
\begin{equation}
 A=0,\quad \psi_s-\psi_o+B=0.
 \label{eq:appCircularLensEquations}
\end{equation}
For their transverse variation, $\theta_s$ is allowed to vary at
fixed $r_s$ before setting $\theta_s=\pi/2$. The angular velocity is
held at its fiducial value throughout this test family; the neighboring
endpoints need not lie on circular geodesics. Define
\begin{equation}
 J_K=\det\begin{pmatrix}
 \partial_l A&\partial_\eta A\\
 \partial_l B&\partial_\eta B
 \end{pmatrix}_{r_s,\theta_s=\pi/2,O,\Omega_K}.
 \label{eq:appABDeterminant}
\end{equation}
These are partial derivatives at fixed endpoint coordinates and branch, not total derivatives along a family satisfying $A=0$. Moving turning
points contribute when the path integrals are differentiated. 

For steady isotropic emission with total source-frame luminosity $L_s$, let $F_{\rm N}=L_s/(4\pi r_o^2)$ be the spherical-dilution reference. In the point-source and asymptotically distant observer limits, the frequency-integrated flux of one regular image is
\begin{equation}
 \left(\frac{F_o}{F_{\rm N}}\right)_{\rm lens}
 =\frac{g^3}
 {2u_s^t\sin\theta_o\sqrt{\eta R_s\Theta_o}\,|J_K|},
 \label{eq:appStandardKerrFlux}
\end{equation}
where $R_s=R(r_s)$ and $\Theta_o=\Theta(\theta_o)=\beta^2$.
This is Eq.~(2.44) of Ref.~\cite{Gralla:2017ufe}, expressed in our
impact parameters. That reference uses $\widehat\lambda=l$ and
$\widehat q=\sqrt\eta$, with the rows ordered as $(B,A)$, so
\begin{equation}
 \left|\det\frac{\partial(B,A)}
 {\partial(\widehat\lambda,\widehat q)}\right|
 =2\sqrt\eta\,|J_K|.
 \label{eq:appLiteratureJacobian}
\end{equation}
Eq.~\eqref{eq:appStandardKerrFlux} does not use a small-spin,
weak-deflection, or near-extremal expansion. It is an exact
geometric-optics expression in terms of the lens integrals and their
derivatives; finding the image still requires solving the branchwise
lens equations. It assumes the point-source limit of a uniformly,
isotropically emitting small sphere. The Kerr spacetime, the chosen
orbit, and the distant observer are held fixed in taking this limit.
The formula applies away from caustics and in a regular coordinate
chart. This flux ratio is distinct from the geometric squared-strain factor
$\mu$.

\subsection{Recovery from the bilocal geodesic operators}
\label{app:KerrBGOEquivalence}
\label{app:KerrCuts}
\label{app:CircularKerrEquivalence}

We now calculate the same flux from the Jacobi map. Begin at a finite
observation event $O$ and label neighboring rays by $(\alpha,\beta)$.
At finite $r_o$ these labels are defined from $(l,\eta)$, rather than
being exact angular coordinates. Their intersections with $r=r_s$ generally occur at different affine
parameters. With $\lambda_o$ fixed, differentiation of the geodesic
flow gives \cite{Serbenta:2021tzv}
\begin{equation}
 \partial_i x_s^\mu
 =W_{XK}^{\mu}{}_{\nu'}(\lambda_s,\lambda_o)
             \partial_i k_o^{\nu'}
  +k_s^\mu\partial_i\lambda_s,
 \label{eq:appEndpointVariation}
\end{equation}
where $(\partial_1,\partial_2)=(\partial_\alpha,\partial_\beta)$.
The primed index refers to $O$. Since this event is fixed, there is
no initial position variation. At fixed affine parameter, the first term is a Jacobi
field with observer-vertex data. The second term locates the
intersection with the source cut.

For the observer's orthonormal screen, write the initial angular
variation as
\begin{equation}
 \begin{aligned}
 & \partial_i k_o^\mu
=\omega_o e_B{}^\mu(O)\mathcal E^B{}_i
    +(\partial_i\ln\omega_o)k_o^\mu,\\
& \mathcal E^B{}_i=e^B{}_{\mu}(O)\partial_i n_o^\mu.
 \end{aligned}
 \label{eq:appAngularVariation}
\end{equation}
The observer four-velocity is fixed. Thus $\mathcal E^B{}_i$ converts
the ray labels to local orthonormal angular increments. Its absolute
determinant is the sky density,
\begin{equation}
 \md\Omega_o^{\rm sky}
 =\mathcal K_o|\md\alpha\,\md\beta|,\quad
 \mathcal K_o=|\det\mathcal E|.
 \label{eq:appSkyMeasure}
\end{equation}
The antipodal change from propagation direction to incoming image
direction does not change this measure. Eq.~\eqref{eq:appNullContractions}
maps the second term of Eq.~\eqref{eq:appAngularVariation} into a
longitudinal displacement. Screen projection therefore removes both
this term and the intersection shift in
Eq.~\eqref{eq:appEndpointVariation}. In compatible endpoint bases,
\begin{equation}
 e^A{}_{\mu}(S)\partial_i x_s^\mu
 =\omega_o\mD_o^A{}_{B}\mathcal E^B{}_i.
 \label{eq:appProjectedLensMap}
\end{equation}
This identifies the differential of the lens map with the transverse
BGO before taking its determinant. The source screen can be changed to the circular source's rest frame
without changing its area.

The intersections lie on the observer's past light cone. Since
$k\cdot\partial_i x$ vanishes at the vertex and is conserved, a
variation on the fixed-radius cut satisfies
\begin{equation}
 \md t_s=k_{\theta,s}\md\theta_s+l\md\phi_s,\quad
 k_{\theta,s}=v_{\theta_s}\sqrt{\Theta_s}.
 \label{eq:appLightConeCut}
\end{equation}
The cut tangents are $V_\theta=(k_\theta,0,1,0)$ and
$V_\phi=(l,0,0,1)$. In Kerr,
$g_{\theta\theta}=\Sigma=r^2+a^2\cos^2\theta$ and
$g_{tt}g_{\phi\phi}-g_{t\phi}^2=-\Delta\sin^2\theta$, giving
\begin{equation}
 \begin{aligned}
 \det[g(V_a,V_b)]
 &=\Sigma(g_{\phi\phi}+2lg_{t\phi}+l^2g_{tt})\\
 &\quad+\Theta(g_{tt}g_{\phi\phi}-g_{t\phi}^2)\\
 &=R\sin^2\theta.
 \end{aligned}
 \label{eq:appSphereGram}
\end{equation}
Projection preserves inner products on $k^\perp$. The physical
transverse source area is consequently
\begin{equation}
 \md\mathscr A_s=\sqrt{R_s}\sin\theta_s
                    |\md\theta_s\,\md\phi_s|.
 \label{eq:appSphereArea}
\end{equation}
This area includes the incidence projection and light-cone time shift;
it is not the area of a constant-$t$ coordinate sphere.

Let
\begin{equation}
 J_r=\det\left[
 \frac{\partial(\cos\theta_s,\phi_s)}{\partial(\alpha,\beta)}
 \right]_{r_s,O}.
 \label{eq:appSphereJacobian}
\end{equation}
Taking the determinant of Eq.~\eqref{eq:appProjectedLensMap} and using
Eqs.~\eqref{eq:appSkyMeasure} and \eqref{eq:appSphereArea} yields
\begin{equation}
 D_{\rm A}^2=\omega_o^2|\det\mD_o|
 =\frac{\sqrt{R_s}|J_r|}{\mathcal K_o}.
 \label{eq:appJacobiLensArea}
\end{equation}
The source-normalized quantity follows by reciprocity,
$\mD_s=-\mD_o^T$, rather than by reversing the frequency factors.
Using the reference area $r_o^2$,
\begin{equation}
 \mu=\frac{r_o^2}{\omega_s^2|\det\mD_s|}
 =\frac{r_o^2g^2\mathcal K_o}{\sqrt{R_s}|J_r|}.
 \label{eq:appFiniteObserver}
\end{equation}
This relation is exact at finite observer radius. The local sky
measure, if needed without an explicit tetrad, is
\begin{equation}
 \mathcal K_o=\frac{\sqrt{\det H}}{\omega_o^2},\quad
 H_{ij}=g_o^{\mu\nu}(\partial_i k_\mu)(\partial_j k_\nu).
 \label{eq:appExactSkyDensity}
\end{equation}
For a static observer with $r_o\to\infty$, the Bardeen labels give
$r_o^2\mathcal K_o\to1$. Hence
\begin{equation}
 \mu=\frac{g^2}{\sqrt{R_s}|J_r|}
 \quad(r_o\to\infty).
 \label{eq:appKerrSphereMagnification}
\end{equation}
For the circular source, the light-cone time shift also enters the
corotating coordinate. At fixed $\Omega_K$,
\begin{equation}
 \md\psi_s=(1-\Omega_Kl)\md\phi_s
              -\Omega_Kk_{\theta,s}\md\theta_s.
 \label{eq:appCorotatingDifferential}
\end{equation}
Define $J_\psi=\det[\partial(\theta_s,\psi_s)/\partial(\alpha,\beta)]_{r_s,O}$.
On the central equatorial ray,
\begin{equation}
 |J_\psi|=(1-\Omega_Kl)|J_r|,\quad
 1-\Omega_Kl>0.
 \label{eq:appCorotatingJacobian}
\end{equation}
The positivity follows from $\omega_s=u_s^t(1-\Omega_Kl)>0$.
Because $t_s=t_o-\Delta t$, omitting the derivative of $\Delta t$
would omit part of this Jacobian.

It remains to express $J_\psi$ through the lens integrals. With
$y=(\theta_s,\psi_s)$ and $c=(l,\eta)$, define
$\boldsymbol f=(A,\psi_s-\psi_o+B)$. Implicit differentiation gives
\begin{equation}
 \frac{\partial y}{\partial c}
 =-\boldsymbol f_{,y}^{-1}\boldsymbol f_{,c},\quad
 \boldsymbol f_{,y}=\begin{pmatrix}
 v_{\theta_s}/\sqrt\eta&0\\
 B_{,\theta_s}&1
 \end{pmatrix}
 \label{eq:appABEndpoint}
\end{equation}
at $\theta_s=\pi/2$. The entry $B_{,\theta_s}$ need not vanish,
but does not enter the determinant. Using the Bardeen coordinates,
\begin{equation}
 \left|\det\frac{\partial(l,\eta)}{\partial(\alpha,\beta)}\right|
 =2|\beta|\sin\theta_o,
 \label{eq:appImpactJacobian}
\end{equation}
and therefore
\begin{equation}
 \begin{aligned}
\left|\det\frac{\partial(\theta_s,\psi_s)}{\partial(l,\eta)}\right|
=\sqrt\eta\,|J_K|,\quad |J_\psi|=2\sin\theta_o\sqrt{\eta\Theta_o}\,|J_K|.
 \end{aligned}
 \label{eq:appABJacobianChain}
\end{equation}
Substitution into Eq.~\eqref{eq:appKerrSphereMagnification}, with
Eq.~\eqref{eq:appCircularRedshift}, gives
\begin{equation}
 \lim_{r_o\to\infty}\frac{r_o^2}{\omega_s^2|\det\mD_s|}
 =\frac{g}
 {2u_s^t\sin\theta_o\sqrt{\eta R_s\Theta_o}\,|J_K|}.
 \label{eq:appBGOIntegralEquivalence}
\end{equation}
All quantities on the right are evaluated in the same distant-observer
limit. For the steady isotropic source specified above,
Eq.~\eqref{eq:appIntegratedFlux} then yields
\begin{equation}
 \left(\frac{F_o}{F_{\rm N}}\right)_{\rm BGO}
 =g^2\mu
 =\left(\frac{F_o}{F_{\rm N}}\right)_{\rm lens}.
 \label{eq:appRecoveredKerrFlux}
\end{equation}
This recovers Eq.~\eqref{eq:appStandardKerrFlux} including its absolute
normalization. The factor $g^2$ converts the matched-source
squared-strain magnification into the integrated flux ratio for this
reference. For a directional spectral source, the corresponding
relation remains
\begin{equation}
 F_{\nu_o}=\frac{g\mu}{r_o^2}
 \left.\mathcal L_{\nu_s}(\tau_s,n_s)\right|_{\nu_s=\nu_o/g}.
 \label{eq:appKerrSpectralFlux}
\end{equation}

The area can also be evaluated on the equatorial cut. Its tangents
$(k_r,1,0,0)$ and $(l,0,0,1)$ have Gram determinant $\eta$, giving
\begin{equation}
 \begin{aligned}
 J_{\rm eq}=\det\left[
 \frac{\partial(r_s,\phi_s)}{\partial(\alpha,\beta)}
 \right]_{\theta_s=\pi/2,O}\!\!,\,\,\,  \sqrt{R_s}|J_r|=\sqrt\eta\,|J_{\rm eq}|.
 \end{aligned}
 \label{eq:appCutIdentity}
\end{equation}
This provides a second coordinate representation of the same
transverse map. The fixed-radius cut requires $R_s>0$. The implicit differentiation
above assumes $\eta>0$, while the Bardeen-to-$(l,\eta)$ conversion
uses $\beta\ne0$ and $\sin\theta_o\ne0$. A grazing cut or a coordinate
degeneracy requires another regular chart and does not by itself
identify a caustic. The derivatives are two dimensional even when
the fiducial source lies on a one-dimensional circular orbit.

In practice, the derivatives are taken at finite fixed $r_o$ before
the asymptotic limit. An additive divergent part of $\Delta t$ may
be subtracted only if it is independent of $(l,\eta)$. Neither the
area equivalence nor the flux comparison regularizes a caustic.
Coherent GW images still require amplitude-level superposition with
retarded emission data, polarization transport, and propagation and
caustic phases; diffraction is needed near a singular geometric map.

\twocolumngrid

 \bibliographystyle{utphys}

\bibliography{refs}

\end{document}